%Paper: hep-th/9505056
%From: ZUCCHINIR@bologna.infn.it
%Date: Wed, 10 May 1995 14:55:47 +0200 (WET-DST)

\magnification=1200
\hoffset=.0cm
\voffset=.0cm
\baselineskip=.55cm plus .55mm minus .55mm
%       This is for referencing
%
\def\ref#1{\lbrack#1\rbrack}
%
%%%%%%%%%%%%%%%%%%%%%%%%%%%%%%%%%%%%%%%%%%%%%%%%%%%%%%%%%%%%%%%%%%%%%%%%%%%%%%%
%
%       Font loading
%
%	The following makes the AmS fonts available
%	(needs the files amssym.def and amssym.tex)
%
\input amssym.def
\input amssym.tex
%
%       Some more fonts:
%
\font\teneusm=eusm10
\font\seveneusm=eusm7
\font\fiveeusm=eusm5
%
%       loads 10, 7 and 5 points size Euler script medium weight
%       (Euler frak (eufm) already loaded by amssym.def)
%
\font\sf=cmss10
\font\ssf=cmss8
%
%       loads 10 and 8 points sans serif
%       (this is a standard plain TeX font)
%
\font\cps=cmcsc10
%
%       loads 10 points capital plus small capital
%       (this is a standard plain TeX font)
%
\newfam\eusmfam
\textfont\eusmfam=\teneusm
\scriptfont\eusmfam=\seveneusm
\scriptscriptfont\eusmfam=\fiveeusm
\def\sans#1{\hbox{\sf #1}}
\def\ssans#1{\hbox{\ssf #1}}

\def\proclaim #1. #2\par{\medbreak{\cps #1.\enspace}{\it #2}\par\medbreak}
%
%        End of font calling
%        Standard abbreviations
%

\def\dtr{\hskip 1pt{\rm det}\hskip 1pt}

\def\ind{\hskip 1pt{\rm ind}\hskip 1pt}
\def\ker{\hskip 1pt{\rm ker}\hskip 1pt}
\def\coker{\hskip 1pt{\rm coker}\hskip 1pt}

\def\dom{\hskip 1pt{\rm dom}\hskip 1pt}

\def\id{\hskip 1pt{\rm id}\hskip 1pt}
\def\tr{\hskip 1pt{\rm tr}\hskip 1pt}
\def\Tr{\hskip 1pt{\rm Tr}\hskip 1pt}
\def\ad{\hskip 1pt{\rm ad}\hskip 1pt}
\def\Ad{\hskip 1pt{\rm Ad}\hskip 1pt}
\def\Lie{\hskip 1pt{\rm Lie}\hskip 1pt}
\def\End{\hskip 1pt{\rm End}\hskip 1pt}

\def\Aut{\hskip 1pt{\rm Aut}\hskip 1pt}
\def\Gau{\hskip 1pt{\rm Gau}\hskip 1pt}

\def\Teich{\hskip 1pt{\rm Teich}\hskip 1pt}

\def\Diff{\hskip 1pt{\rm Diff}\hskip 1pt}


\hrule\vskip.4cm
\hbox to 16.5 truecm{May 1995   \hfil DFUB 95--3}
\hbox to 16.5 truecm{Version 1  \hfil hep-th/9505056}
\vskip.4cm\hrule\vskip1.5cm
\centerline{\bf EXTRINSIC HERMITIAN GEOMETRY OF FUNCTIONAL}
\centerline{\bf DETERMINANTS FOR VECTOR SUBBUNDLES}
\centerline{\bf AND THE DRINFELD--SOKOLOV GHOST SYSTEM}
\vskip1cm
\centerline{by}
\vskip.5cm
\centerline{\bf Roberto Zucchini}
\centerline{\it Dipartimento di Fisica, Universit\`a degli Studi di Bologna}
\centerline{\it V. Irnerio 46, I-40126 Bologna, Italy}
\vskip1.5cm\hrule\vskip.9cm
\centerline{\bf Abstract}
\vskip.3cm
\par\noindent
In this paper, a novel method is presented for the study of the dependence
of the functional determinant of the Laplace operator associated to a
subbundle $F$ of a hermitian holomorphic vector bundle $E$ over a Riemann
surface $\Sigma$ on the hermitian structure $(h,H)$ of $E$. The generalized
Weyl anomaly of the effective action is computed and found to be
expressible in terms of a suitable generalization of the Liouville and
Donaldson actions. The general techniques worked out are then applied
to the study of a specific model, the Drinfeld--Sokolov (DS) ghost system
arising in $W$--gravity. The expression of generalized Weyl anomaly of the
DS ghost effective action is found. It is shown that, by a specific
choice of the fiber metric $H_h$ depending on the base metric $h$,
the effective action reduces into that of a conformal field theory. Its
central charge is computed and found to agree with that obtained by the
methods of hamiltonian reduction and conformal field theory. The
DS holomorphic gauge group and the DS moduli space are defined
and their dimensions are computed.
\vfill\eject
\item{1.} {\bf Introduction}
\vskip.4cm
\par
In the last thirty years, a large body of physical literature has been
devoted to the study of functional determinants in connection with
quantum gravity, gauge theory and, more recently, string theory. Several
methods for their computation have been developed such as zeta function
regularization \ref{1--5}, proper time regularization \ref{6} and
Fujikawa's method \ref{7} to mention only the most frequently used.
All these approaches analyze the dependence of the determinants on
the relevant background fields and
employ in a crucial manner the Seeley--De Witt
coefficients of the associated heat kernels \ref{8--10}.

In this paper, a novel method is presented for the analysis of
functional determinants of Laplace operators associated to
a subbundle of a holomorphic vector bundle on a Riemann surface
from an extrinsic point of view. The general techniques worked out
are then applied to the study of a specific model, the Drinfeld--Sokolov
ghost system arising in $W$--gravity. The results obtained in this
way are interesting both as illustration of the general formalism and
for its applications to $W$--strings.

The problem tackled in the first part of this paper can be stated
as follows. Consider a holomorphic vector bundle $E$ on a Riemann
surface $\Sigma$ and a subbundle $F$ of $E$. A hermitian structure
$(h,H)$ on $E$ is a pair consisting of a hermitian metric $h$ on
$\Sigma$ and a hermitian fiber metric $H$. When $E$ is equipped
with a hermitian structure $(h,H)$, a hermitian structure
$(h,H_F)$ is induced on the subbundle $F$.
This allows to construct the Laplace operator
$\Delta_{w,F;h,H}=\bar\partial^\star_{w,F;h,H}\bar\partial_{w,F;h,H}$
associated to the Cauchy--Riemann operator $\bar\partial_{w,F}$
acting on $F$--valued conformal fields of weight $w$.
Using proper time regularization, one may then define the determinant
$\dtr'_\epsilon\Delta_{w,F;h,H}$, where $\epsilon$ is the proper time
ultraviolet cut--off.
Two different approaches to the study of such determinant can be
envisaged. In the `intrinsic' approach, one considers $F$ as a holomorphic
vector bundle on its own right equipped with the induced hermitian structure
$(h,H_F)$. The problem is then reduced to the standard one of studying
$\dtr'_\epsilon\Delta_{w,E;h,H}$ for a holomorphic vector bundle $E$
endowed with a hermitian structure $(h,H)$ \ref{11}. This approach has the
drawback that all results are expressed in terms of $H_F$, which depends
in a complicated way on $H$, while, in certain physical applications,
one would like to express the results directly in terms of $H$. Hence,
an `extrinsic' approach capable of computing the dependence of
$\dtr'_\epsilon\Delta_{w,F;h,H}$ on the hermitian structure $(h,H)$
of $E$ would be desirable.

In sect. 2, $\dtr'_\epsilon\Delta_{w,F;h,H}$ is studied as a functional
of $(h,H)$ in the framework of the Liouville--Donaldson parametrization of
the family of hermitian structures \ref{12}. The expressions obtained
involve the $H$--hermitian fiber projector $\varpi_{F;H}$ of $E$
onto $F$, which is a local functional of $H$. In sect. 3, the class of
special holomorphic structures of the smooth vector bundle $E$, for which
the smooth subbundle $F$ is holomorphic, is characterized in the framework
of the Beltrami--Koszul parametrization of the holomorphic structures
\ref{12}. Further, it is shown that the special subgroup of the automorphism
group of $E$ preserving $F$ preserves such class of holomorphic structures
and is the symmetry group under which $\dtr'_\epsilon\Delta_{w,E;h,H}$
is invariant.

In the second part of the paper, the results outlined above are applied
to the study of the renormalized effective action of the
Drinfeld--Sokolov ghost system in $W$--gravity \ref{13--15}.
Let us briefly recall the formulation of the model.
Let $G$ be a simple complex Lie group and let $S$ be an
$SL(2,\Bbb C)$ subgroup of $G$ invariant under the compact conjugation
$\dagger$ of $G$. To these algebraic data, there is associated a
halfinteger grading of $\goth g$ and a certain bilinear form $\chi$ on
$\goth g$ \ref{14}. On a Riemann surface $\Sigma$ with a spinor structure
$k^{\otimes {1\over 2}}$, one can further associate to the pair $(G,S)$
a holomorphic $G$--valued cocycle defining holomorphic principal
$G$--bundle, the Drinfeld--Sokolov (DS) bundle $L$ \ref{16}.
$\Ad L$ is then a holomorphic vector bundle. If $\goth x$ is a maximal
negative graded subalgebra of $\goth g$ isotropic with respect to $\chi$,
then the $\goth x$--valued sections of $\Ad L$ span a holomorphic subbundle
$\Ad L_{\goth x}$ of $\Ad L$.

$W$ gravity may be formulated as a gauge theory based on the smooth
principal $G$ bundle underlying the DS bundle $L$. The gauge fields are
$\goth x$--valued sections of $\bar k\otimes\Ad L$. The gauge group,
the DS gauge group $\Gau_{cDS}$, consists of the $\exp\goth x$--valued
gauge transformations. Fixing such gauge symmetry yields the DS ghosts
$\beta$, $\gamma$ as Fadeev--Popov ghosts. Here, $\beta$ is an
anticommuting section of $k\otimes\Ad L$ valued in $\goth g/\goth x^\perp$,
where $\goth x^\perp$ is the orthogonal complement of
$\goth x$ with respect to the Cartan Killing form $\tr_{\rm ad}$
of $\goth g$. $\gamma$ is an anticommuting $\goth x$--valued section
of $\Ad L$. The action is given by
$$
S_{DS}(\beta,\beta^\dagger,\gamma,\gamma^\dagger)
={1\over\pi}\int_\Sigma d^2z \tr_{\rm ad}\big(\beta\bar\partial\gamma)+
{\rm c.~c.}.
\eqno(1.1)
$$
The quantum effective action is thus related to the determinant
$\dtr'_\epsilon\Delta_{0,\Ad L_{\goth x};h,\Ad H}$ relative to the
holomorphic vector subbundle $\Ad L_{\goth x}$ of $\Ad L$, where $H$
is a hermitian metric on $L$.
In sect. 4, the effective action is studied by
means of the general methods developed in sects. 3 and 4. It is shown
that the metric $h$ can be lifted to a metric $H_h$ on $L$ depending only
$h$. Setting $H=H_h$ in the determinant, one finds that the resulting
renormalized effective action is that of a conformal field theory
perhaps perturbed by a term of the form $\int\sqrt hR_h{}^2$ as
in the model considered in refs. \ref{18--19}. Its central charge
is computed and found to agree with that obtained by the methods of
hamiltonian reduction and conformal field theory \ref{14}. The
dimensions of the spaces of $\beta$-- and $\gamma$--zero modes
and the index of the ghost kinetic operator are also computed.
Finally, the relevant classes of special holomorphic structures
and special automorphisms of $\Ad L$ are defined and studied. A notion
of stability for holomorphic structures is introduced. The holomorphic
DS gauge group and of the DS moduli space of stable holomorphic structures
are then defined and their dimensions computed. It must be emphasized that
the above is the DS moduli space and is distinct from
the $W$--moduli space introduced by Hitchin in ref. \ref{20}
and later identified with the moduli space of quantum $W$--gravity
in ref. \ref{21}.
\vskip.4cm
\item{2.} {\bf The determinant of} $\Delta^\sharp_{w,F;h,H}$
\vskip.4cm
\par
In what follows, $E$ is a holomorphic vector bundle of rank $r_E$ over a
compact connected Riemann surface $\Sigma$ of genus $\ell$.
$k^{\otimes{1\over 2}}$ is
a fixed tensor square root of the canonical line bundle $k$ of $\Sigma$,
{\it i. e.} a spinor structure in physical parlance. $F$ is a holomorphic
vector subbundle of $E$ of rank $r_F>0$. See ref. \ref{17} for basic
background.

Let $w,~\bar w\in\Bbb Z/2$. Denote by ${\cal S}_{w,\bar w}$ the complex
vector space of smooth sections of the complex line bundle $k^{\otimes w}
\otimes\bar k^{\otimes \bar w}$. The elements of ${\cal S}_{w,\bar w}$ are
ordinary conformal fields of weights $w,~\bar w$. If $V$ is a smooth vector
bundle related to either $E$ or $F$, denote by ${\cal S}_{w,\bar w,V}$ the
complex vector space of smooth sections of the complex vector bundle
$k^{\otimes w}\otimes\bar k^{\otimes \bar w}\otimes V$. The elements of
${\cal S}_{w,\bar w,V}$ are generalized vector valued conformal fields of
weights $w,~\bar w$.

A hermitian structure $(h,H)$ on $E$ consists of a hermitian metric
$h$ on the base $\Sigma$ and a hermitian fiber metric $H$,
{\it i. e.} a section $h$ of $k\otimes\bar k$ such that $h>0$ and
a section $H$ of $E\otimes \bar E$ such that $H=H^\dagger>0$.
To any hermitian structure $(h,H)$ on $E$ there is associated a
Hilbert inner product on ${\cal S}_{w,\bar w,F}$ by
$$
\langle \phi,\psi\rangle_{w,\bar w,F;h,H}
=\int_\Sigma d^2 z h^{1-w-\bar w}\phi^\dagger H^{-1}\psi,
\quad \phi,~\psi\in{\cal S}_{w,\bar w,F}.
\eqno(2.1)
$$
By completing ${\cal S}_{w,\bar w,F}$ with respect to the corresponding
norm, one obtains a complex Hilbert space ${\cal H}_{w,\bar w,F;h,H}$
containing ${\cal S}_{w,\bar w,F}$ as a dense subspace.

The Cauchy--Riemann operator $\bar\partial_{w,F}$ is the linear operator
from ${\cal S}_{w,0,F}$ to ${\cal S}_{w,1,F}$ locally given by
$\bar\partial_{w,F}=\bar\partial$ on ${\cal S}_{w,0,F}$.
$\bar\partial_{w,F}$
can be extended to a linear operator from a dense subspace of
${\cal H}_{w,0,F;h,H}$ containing ${\cal S}_{w,0,F}$ into
${\cal H}_{w,1,F;h,H}$. Its adjoint $\bar\partial^\star_{w,F;h,H}$
is a linear operator from a dense subspace of
${\cal H}_{w,1,F;h,H}$ containing ${\cal S}_{w,1,F}$ into
${\cal H}_{w,0,F;h,H}$. Using $\bar\partial_{w,F;h,H}$ and
$\bar\partial^\star_{w,F;h,H}$, one can define the Laplace operators
$$
\Delta_{w,F;h,H}=\bar\partial^\star_{w,F;h,H}\bar\partial_{w,F;h,H},
\eqno(2.2)
$$
$$
\Delta^\vee_{w,F;h,H}=\bar\partial_{w,F;h,H}\bar\partial^\star_{w,F;h,H}.
\eqno(2.3)
$$
$\Delta_{w,F;h,H}$ is a linear operator from a dense subspace of
${\cal H}_{w,0,F;h,H}$ containing ${\cal S}_{w,0,F}$ into
${\cal H}_{w,0,F;h,H}$. $\Delta^\vee_{w,F;h,H}$ is a linear operator from a
dense subspace of ${\cal H}_{w,1,F;h,H}$ containing ${\cal S}_{w,1,F}$ into
${\cal H}_{w,1,F;h,H}$. $\Delta_{w,F;h,H}$ and $\Delta^\vee_{w,F;h,H}$ are
essentially self--adjoint unbounded elliptic linear differential operators
with a discrete non negative spectrum of finite multiplicity.
Furthermore, $\Delta_{w,F;h,H}$ and $\Delta^\vee_{w,F;h,H}$ have the same
spectrum and their non zero eigenvalues have the same multiplicity.

One has that $\ker\bar\partial_{w,F;h,H}=\ker\Delta_{w,F;h,H}
\cong\ker\bar\partial_{w,F}$ and $\ker\bar\partial^\star_{w,F;h,H}=
\ker\Delta^\vee_{w,F;h,H}$ $\cong\coker\bar\partial_{w,F}$. The difference
$\ind\bar\partial_{w,F}=\dim\ker\bar\partial_{w,F}-
\dim\coker\bar\partial_{w,F}$ is the Atiyah--Singer index of
$\bar\partial_{w,F}$ and is a topological invariant, {\it i. e.}
it is independent from the background holomorphic structure of $E$ and
from the hermitian structure $(h,H)$. One has thus
$$
\ind\bar\partial_{w,F}
=\dim\ker\bar\partial_{w,F;h,H}-\dim\ker\bar\partial^\star_{w,F;h,H}
=\dim\ker\Delta_{w,F;h,H}-\dim\ker\Delta^\vee_{w,F;h,H}.
\eqno(2.4)
$$

In field theory, the objects of main interest are the functional
determinants of $\Delta_{w,F;h,H}$ and $\Delta^\vee_{w,F;h,H}$.
In this paper, these will be defined by the proper time method
\ref{6}. In such approach, the zero eigenvalues are excluded in order
to get a non trivial result. Further, since the spectrum of the
operators considered is not bounded above, it is necessary to introduce an
ultraviolet cut--off $1/\epsilon$ with $\epsilon>0$. One thus uses the
standard notation $\dtr'_\epsilon$ to denote the cut--off determinant
with the zero eigenvalues removed. Since the non zero spectra of
$\Delta_{w,F;h,H}$ and $\Delta^\vee_{w,F;h,H}$ are identical, one knows
a priori that
$$
\dtr'_\epsilon\Delta_{w,F;h,H}=\dtr'_\epsilon\Delta^\vee_{w,F;h,H}.
\eqno(2.5)
$$
It is thus convenient to denote by $\Delta^\sharp_{w,F;h,H}$ either
$\Delta_{w,F;h,H}$ or $\Delta^\vee_{w,F;h,H}$.
Following \ref{6}, one has
$$
\ln\dtr'_\epsilon\Delta^\sharp_{w,F;h,H}
=-\int_\epsilon^\infty{dt\over t}
\Big[\Tr\big(\exp(-t\Delta^\sharp_{w,F;h,H})\big)-d^\sharp_{w,F}\Big],
\eqno(2.6)$$
where $d^\sharp_{w,F}=\dim\ker\Delta^\sharp_{w,F;h,H}$.
Using the small $t$ expansion of the diagonal part of the heat kernel
$\exp(-t\Delta^\sharp_{w,F;h,H})$ of $\Delta^\sharp_{w,F;h,H}$, one
can compute the terms of $\ln\dtr'_\epsilon\Delta^\sharp_{w,F;h,H}$,
which diverge as $\epsilon\rightarrow 0$.

As well--known, it is very difficult to compute
$\dtr'_\epsilon\Delta^\sharp_{w,F;h,H}$ as a functional of $(h,H)$ directly
from $(2.6)$. It is instead relatively easier to compute the variation
of $\dtr'_\epsilon\Delta^\sharp_{w,F;h,H}$ with respect to $(h,H)$.
To this end, one introduces the differential complex
$(\delta,\Omega^*_{\goth H_E})$ where $\delta$ is the differential operator
on the infinite dimensional manifold $\goth H_E$ of hermitian structures
$(h,H)$ on $E$ satisfying $\delta^2=0$ and $\Omega^*_{\goth H_E}$ is the
corresponding exterior algebra. The cohomology ring
$H^*(\delta,\Omega^*_{\goth H_E})$ is trivial since $\goth H_E$ is
contractible.

The variation of the Hilbert space structure defined by $(2.1)$ is
given by an expression of the form
$$\delta\langle\phi,\psi\rangle_{w,\bar w,F;h,H}
=\langle\phi,Q_{w,\bar w,F;h,H}\psi\rangle_{w,\bar w,F;h,H},
\quad \phi,~\psi\in{\cal S}_{w,\bar w,F}.
\eqno(2.7)
$$
Here, $Q_{w,\bar w,F;h,H}$ is an element of the tensor product
${\cal S}_{0,0,\End F}\otimes\Omega^1_{\goth H_E}$, where $\End F$ is
the endomorphism bundle of $F$.
$Q_{w,\bar w,F;h,H}$ acts as multiplicative operator
in ${\cal H}_{w,\bar w,F;h,H}$ valued in $\Omega^1_{\goth H_E}$ and, as
such, it is bounded and self--adjoint.

One can show that there are bases
$\{\omega_{w,F;h,H;i}|i=1,\ldots,d_{w,F}\}$ and
$\{\omega^\vee_{w,F;h,H;i}|i=1,\ldots,d^\vee_{w,F}\}$
of $\ker\Delta_{w,F;h,H}$ and $\ker\Delta^\vee_{w,F;h,H}$,
respectively, such that
$$
\eqalignno{
\delta\omega_{w,F;h,H;i}=&~0,\quad i=1,\ldots,d_{w,F}
&(2.8),\cr
\delta\omega^\vee_{w,F;h,H;i}=&-Q_{w,1,F;h,H}\omega^\vee_{w,F;h,H;i},
\quad i=1,\ldots,d^\vee_{w,F}
&(2.9).\cr}
$$
The Gram matrices of these bases are
$$
\eqalignno{
M_{w,F;h,H}(\omega)_{i,j}
=&~\langle\omega_{w,F;h,H;i},\omega_{w,F;h,H;j}\rangle_{w,0,F;h,H},
\quad i,~j=1,\ldots,d_{w,F},
&(2.10)\cr
M_{w,F;h,H}(\omega^\vee)_{i,j}
=&~\langle\omega^\vee_{w,F;h,H;i},\omega^\vee_{w,F;h,H;j}\rangle_{w,1,F;h,H},
\quad i,~j=1,\ldots,d^\vee_{w,F}.
&(2.11)\cr}
$$

A standard analysis shows that
$$
\eqalignno{
&\delta\ln\dtr'_\epsilon\Delta^\sharp_{w,F;h,H}
=\delta\ln\det M_{w,F;h,H}(\omega)+\delta\ln\det M_{w,F;h,H}(\omega^\vee)
&\cr
&-\Tr\big(Q_{w,0,F;h,H}\exp(-\epsilon\Delta_{w,F;h,H})\big)
+\Tr\big(Q_{w,1,F;h,H}\exp(-\epsilon\Delta^\vee_{w,F;h,H})\big).
&(2.12)\cr}
$$
The last two traces can be dealt with by using the
small $t$ expansion of the diagonal part of the heat kernel
$\exp(-t\Delta^\sharp_{w,F;h,H})$ of $\Delta^\sharp_{w,F;h,H}$.
In principle, this relation can be integrated and yield an expression
for $\dtr'_\epsilon\Delta^\sharp_{w,F;h,H}$ up to a constant. The part
of the constant that diverges in the ultraviolet limit
$\epsilon\rightarrow 0$ can be computed from $(2.6)$.

The above method for computing functional determinants, and other
methods as well, exploit heat kernel techniques in an essential way.
The elliptic operators $\Delta$ considered here act on a suitable space
of sections of some smooth vector bundle $V$ and are given locally
by an expression of the form
$$
\Delta=-h^{-1}\big(1\bar\partial\partial+\sigma\bar\partial
+\sigma^*\partial+\tau\big),
\eqno(2.13)
$$
where $\sigma$, $\sigma^*$ and $\tau$ are certain smooth matrix valued
functions. A standard calculation on the same line as those described in
ref. \ref{6} yields for the diagonal part of the heat kernel $\exp(-t\Delta)$
of $\Delta$ the local expression
$$
\exp(-t\Delta)_{\rm diag}
={1\over\pi t}h1-{1\over 6\pi}\bar\partial\partial\ln h1
-{1\over 2\pi}\big(\partial\sigma^*+\bar\partial\sigma
+\sigma^*\sigma+\sigma\sigma^*-2\tau\big)+O(t).
\eqno(2.14)
$$
By covariance, $\exp(-\Delta)_{\rm diag}$ must belong to
${\cal S}_{1,1,\End V}$.

One must now proceed to the implementation of the methods described
above. As explained in the introduction, instead
of considering the hermitian structure induced by $(h,H)$ on $F$ and
carrying out the calculation intrinsically, one is interested in expressing
the determinant directly in terms of $(h,H)$. For such reason, one
introduces the orthogonal projector $\varpi_{w,\bar w,F;h,H}$ of
${\cal H}_{w,\bar w,E;h,H}$ onto ${\cal H}_{w,\bar w,F;h,H}$.
$\varpi_{w,\bar w,F;h,H}$ is a bounded self--adjoint multiplicative
operator corresponding to a smooth section $\varpi_{F;H}$ of
$F\otimes E^\vee$ independent from $w$, $\bar w$ and $h$, where
$E^\vee$ is the dual vector bundle of $E$. As $F$ is a subbundle of
$E$, $\varpi_{F;H}$ is also an element of ${\cal S}_{0,0,\End E}$. The
fact that $\varpi_{w,\bar w,F;h,H}$ is an orthogonal projector
in ${\cal H}_{w,\bar w,E;h,H}$ implies that
$$
\varpi_{F;H}{}^2=\varpi_{F;H}
\eqno(2.15)
$$
$$
H\varpi_{F;H}{}^\dagger H^{-1}=\varpi_{F;H}.
\eqno(2.16)
$$

If $\phi\in{\cal S}_{w,0,F}$, then $\bar\partial_{w,F}\phi\in
{\cal S}_{w,1,F}$, so that one has at the same time that
$\varpi_{F;H}\phi=\phi$ and $\varpi_{F;H}\bar\partial_{w,F}\phi=
\bar\partial_{w,F}\phi$. From this remark and $(2.16)$, the following
equivalent relations follow
$$
\bar\partial\varpi_{F;H}\varpi_{F;H}=0,\quad
\varpi_{F;H}\partial_H\varpi_{F;H}=0,
\eqno(2.17)
$$
where $\partial_H=\partial-\ad(\partial HH^{-1})$ is the covariant derivative
on $\End E$ associated to the metric connection $\partial HH^{-1}$
\footnote{${}^1$}{By convention, each differential operator acts on the object
immediately at its right.}.
Projectors $\varpi_{F;H}$ satisfying $(2.17)$ were introduced
earlier in the mathematical literature in the analysis of
Hermitian--Einstein and Higgs bundles \ref{22--23}.

If $\phi\in{\cal S}_{w,0,F}$, then $\varpi_{F;H}\phi=\phi$, so that
$\varpi_{F;H}\phi$ is independent from $H$.  This implies the relation
$$
\delta\varpi_{F;H}\varpi_{F;H}=0.
\eqno(2.18)
$$
By combining $(2.16)$ and $(2.18)$, one obtains
$$
\delta\varpi_{F;H}=-\varpi_{F;H}\delta HH^{-1}(1-\varpi_{F;H}).
\eqno(2.19)
$$
This identity is of crucial importance. It is a functional
differential equation constraining the dependence of $\varpi_{F;H}$
on $H$ and shows that $\varpi_{F;H}$ is a local functional of $H$.

By direct calculation, one finds that the differential
operators $\bar\partial_{w,F;h,H}$, $\bar\partial^\star_{w,F;h,H}$,
$\Delta_{w,F;h,H}$ and $\Delta^\vee_{w,F;h,H}$ defined earlier have
the following local expressions
$$\eqalignno{
\bar\partial_{w,F;h,H}=&\hskip 1pt
1\bar\partial, \quad{\rm on}~{\cal S}_{w,0,F},
\vphantom{\Big[}&(2.20)\cr
\bar\partial^\star_{w,F;h,H}
=&-h^{-1}\big(1\partial-w\partial\ln h 1-\partial HH^{-1}
-\partial_H\varpi_{F;H}\big), \quad{\rm on}~{\cal S}_{w,1,F},
\vphantom{\Big[}&(2.21)\cr
\Delta_{w,F;h,H}=&-h^{-1}\big[1\bar\partial\partial
-(w\partial\ln h 1+\partial HH^{-1}+\partial_H\varpi_{F;H})\bar\partial
\big], \quad{\rm on}~{\cal S}_{w,0,F},
\vphantom{\Big[}&(2.22)\cr
\Delta^\vee_{w,F;h,H}=&-h^{-1}\big[1\bar\partial\partial
-\bar\partial\ln h1\partial
-(w\partial\ln h 1+\partial HH^{-1}+\partial_H\varpi_{F;H})\bar\partial
-w(f_h
\vphantom{\Big[}&\cr
-\bar\partial\ln h\partial\ln h&)1
+\bar\partial\ln h(\partial HH^{-1}+\partial_H\varpi_{F;H})
-F_H-\bar\partial\partial_H\varpi_{F;H}\big], \quad{\rm on}~{\cal S}_{w,1,F},
{}~~~~\vphantom{\Big[}&(2.23)\cr}
$$
where $f_h=\bar\partial\partial\ln h$
and $F_H=\bar\partial(\partial HH^{-1})$ are the curvatures of $h$ and $H$.
One can easily check that all these operators map $F$--valued conformal
fields into $F$--valued conformal fields.

By $(2.22)$ and $(2.23)$,
the operators $\Delta_{w,F;h,H}$ and $\Delta^\vee_{w,F;h,H}$ are of the
form $(2.13)$. This allows one to obtain the diagonal part of the
corresponding heat kernels by applying $(2.14)$. The resulting local
expressions are
$$
\exp(-t\Delta_{w,F;h,H})_{\rm diag}
={1\over\pi t}h1+{3w-1\over 6\pi}f_h1
+{1\over 2\pi}(F_H+\bar\partial\partial_H\varpi_{F;H})+O(t),
\eqno(2.24)
$$
$$
\exp(-t\Delta^\vee_{w,F;h,H})_{\rm diag}
={1\over\pi t}h1+{2-3w\over 6\pi}f_h1
-{1\over 2\pi}(F_H+\bar\partial\partial_H\varpi_{F;H})+O(t).
\eqno(2.25)
$$
Using $(2.17)$, it can be verified that $\exp(-t\Delta_{w,F;h,H})_{\rm diag}$
and $\exp(-t\Delta^\vee_{w,F;h,H})_{\rm diag}$, as given by
$(2.24)$ and $(2.25)$, belong to ${\cal S}_{1,1,\End F}$, as expected
on general grounds. As they preserve $F$, one has that
$\Tr\big(\exp(-t\Delta^\sharp_{w,F;h,H})Q\big)=\int_\Sigma d^2z
\tr\big(\varpi_{F;H}\exp(-t\Delta^\sharp_{w,F;h,H})_{\rm diag}Q\big)$
for any bounded self--adjoint multiplicative operator $Q$ in the
appropriate Hilbert space corresponding to some element $Q$ of
${\cal S}_{0,0,\End F}$, $\tr$ denoting the ordinary fiber
trace of $\End E$.

Finally one needs to compute the operators $Q_{w,\bar w,F;h,H}$ defined
in $(2.7)$. By varying $(2.1)$, one finds the following local expression
$$
Q_{w,\bar w,F;h,H}=(1-w-\bar w)\delta\ln h
-\delta HH^{-1}+[\delta HH^{-1},\varpi_{F;H}],
\quad{\rm on}~{\cal S}_{w,\bar w,F}.
\eqno(2.26)
$$
One can check that this operator preserves $F$.

Now, one has all the elements required for the study of
$\dtr'_\epsilon\Delta^\sharp_{w,F;h,H}$. In view of physical
applications, it turns out to be more convenient to consider a closely
related object, the unrenormalized effective action
$I^{\rm bare}_{w,F}(h,H;\epsilon)$ defined by
$$
I^{\rm bare}_{w,F}(h,H;\epsilon)
=\ln\bigg[{\dtr'_\epsilon\Delta^\sharp_{w,F;h,H}\over
\det M_{w,F;h,H}(\omega)\det M_{w,F;h,H}(\omega^\vee)}\bigg].
\eqno(2.27)
$$

Let us find the terms of $I^{\rm bare}_{w,F}(h,H;\epsilon)$, which are
singular in the limit $\epsilon\rightarrow 0$ corresponding to the
removal of the cut--off. By plugging either one of $(2.24)$ and $(2.25)$
into $(2.6)$, one gets
$$
I^{\rm bare}_{w,F}(h,H;\epsilon)
=-{r_F\over\pi\epsilon}\int_\Sigma d^2zh+K_{w,F}\ln\epsilon+O(1),
\eqno(2.28)
$$
where
$$\eqalignno{
K_{w,F}=&~{(3w-1)r_F\over 6\pi}\int_\Sigma d^2zf_h
+{1\over 2\pi}\int_\Sigma d^2z
\tr\big((F_H+\bar\partial\partial_H\varpi_{F;H})\varpi_{F;H}\big)-d_{w,F},
{}~~~&\cr
=&~{(2-3w)r_F\over 6\pi}\int_\Sigma d^2zf_h
-{1\over 2\pi}\int_\Sigma d^2z
\tr\big((F_H+\bar\partial\partial_H\varpi_{F;H})\varpi_{F;H}\big)-d^\vee_{w,F}.
{}~~~&(2.29)\cr}
$$
$K_{w,F}$ is independent from $(h,H)$. This can be verified using the
variational relations
$$
\delta f_h=\bar\partial\partial\delta\ln h,
\eqno(2.30)
$$
$$
\delta\tr\big((F_H+\bar\partial\partial_H\varpi_{F;H})\varpi_{F;H}\big)
=\bar\partial\partial\tr\big(\delta HH^{-1}\varpi_{F;H}\big),
\eqno(2.31)
$$
the second of which follows from repeated applications of $(2.17)$ and
$(2.19)$. By comparing the two expressions of $K_{w,F}$, and recalling that
$\ind\bar\partial_{w,F}=\dim\ker\bar\partial_{w,F}-\dim
\coker\bar\partial_{w,F}$, one obtains the index relation
$$
\hfil
\ind\bar\partial_{w,F}
={(2w-1)r_F\over 2\pi}\int_\Sigma d^2zf_h
+{1\over\pi}\int_\Sigma d^2z
\tr\big((F_H+\bar\partial\partial_H\varpi_{F;H})\varpi_{F;H}\big).
\eqno(2.32)
$$
Hence, the right hand side of $(2.32)$ is a topological invariant.
Independence from $(h,H)$ follows from $(2.30)-(2.31)$. Independence
from the background holomorphic structure of $E$ can be verified
employing the Beltrami--Koszul parametrization of holomorphic
structures described in sect. 3.
Indeed, the above integrals are respectively, up to factors,
the Gauss--Bonnet invariant and the Chern-Weil invariant of $F$
\ref{23}.

Next, let us compute $\delta I^{\rm bare}_{w,F}(h,H;\epsilon)$.
By using $(2.12)$, $(2.24)-(2.25)$ and $(2.26)$, one finds
$$
\delta I^{\rm bare}_{w,F}(h,H;\epsilon)
=-{r_F\over\pi\epsilon}\int_\Sigma d^2z\delta h
+A^0_{w,F}(h,H)+O(\epsilon),
\eqno(2.33)
$$
where
$$
\eqalignno{
&A^0_{w,F}(h,H)={(6w^2-6w+1)r_F\over 6\pi}\int_\Sigma d^2z\delta\ln hf_h
&\cr
&+{(2w-1)\over 2\pi}\int_\Sigma d^2z\Big[\delta\ln h
\tr\big((F_H+\bar\partial\partial_H\varpi_{F;H})\varpi_{F;H}\big)
+\tr\big(\delta HH^{-1}\varpi_{F;H}\big)f_h\Big]&\cr
&+{1\over\pi}\int_\Sigma d^2z
\tr\big(\delta
HH^{-1}(F_H+\bar\partial\partial_H\varpi_{F;H})\varpi_{F;H}\big).
&(2.34)\cr}
$$
One may verify that $A^0_{w,F}(h,H)$ is an exact element of
$\Omega^1_{\goth H_E}$. It clearly must be so, since
$\delta I^{\rm bare}_{w,F}(h,H;\epsilon)$ is obviously
exact in $\Omega^1_{\goth H_E}$. The verification relies on $(2.17)$ and
$(2.19)$ and the fact that $H^1(\delta,\Omega^*_{\goth H_E})=0$. Remarkably,
all three terms in the right hand side of $(2.34)$ are
separately exact.

It is possible to integrate $(2.33)$. The integration is carried out along
a functional path in $\goth H_E$ joining a fiducial reference hermitian
structure $(\hat h,\hat H)$ to the hermitian structure considered.
The choice of the path is immaterial because of the
exactness of the $1$--form of $\Omega^1_{\goth H_E}$ integrated. By combining
$(2.28)$ and $(2.33)$, one obtains
$$
I^{\rm bare}_{w,F}(h,H;\epsilon)
=-{r_F\over\pi\epsilon}\int_\Sigma d^2zh+K_{w,F}\ln\epsilon
+S_{w,F}(h,H;\hat h,\hat H)+s_{w,F}(\hat h,\hat H)+O(\epsilon).
\eqno(2.35)
$$
Here, $s_{w,F}(\hat h,\hat H)$ is a finite functional of
$(\hat h,\hat H)$ only and $S_{w,F}(h,H;\hat h,\hat H)$ is formally
given by
$$
S_{w,F}(h,H;\hat h,\hat H)
=\int_{(\hat h,\hat H)}^{(h,H)}A^0_{w,F}(h',H').
\eqno(2.36)
$$
To perform the functional integration, one introduces the Liouville
field $\phi$ of $h$ relative to $\hat h$ and the Donaldson field $\Phi$
of $H$ relative to $\hat H$ \ref{12}. Recall that $\phi$ and $\Phi$
are elements of ${\cal S}_{0,0}$ and ${\cal S}_{0,0,\End E}$,
respectively, such that
$$
h=\exp\phi\hat h,
\eqno(2.37)
$$
$$
\bar\phi=\phi.
\eqno(2.38)
$$
$$
H=\exp\Phi\hat H,
\eqno(2.39)
$$
$$
H\Phi^\dagger H^{-1}=\Phi.
\eqno(2.40)
$$
Using $(2.19)$, it is straightforward to show that $\varpi_{F;H}$ has
a local Taylor expansion in $\Phi$ of the form
$$
\varpi_{F;H}=\sum_{r=0}^\infty{1\over r!}\varpi^{(r)}_{F;\hat H}(\Phi),
\eqno(2.41)
$$
where, for each $r\geq 0$, $\varpi^{(r)}_{F;\hat H}(\Phi)$ is an element of
${\cal S}_{0,0,\End E}$ and a homogeneous degree $r$ polynomial in
$\Phi$:
$$
\eqalignno{
\varpi^{(0)}_{F;\hat H}(\Phi)
=&~\varpi_{F;\hat H},
&\cr
\varpi^{(1)}_{F;\hat H}(\Phi)=&~
-\varpi_{F;\hat H}\Phi(1-\varpi_{F;\hat H}),
&\cr
\varpi^{(2)}_{F;\hat H}(\Phi)=
&~\varpi_{F;\hat H}\Phi(1-2\varpi_{F;\hat H})\Phi(1-\varpi_{F;\hat H}),
&\cr
\varpi^{(3)}_{F;\hat H}(\Phi)=
&~\varpi_{F;\hat H}
\big[\Phi(3\varpi_{F;\hat H}-1)\Phi(1-\varpi_{F;\hat H})\Phi
+\Phi(2-3\varpi_{F;\hat H})\Phi\varpi_{F;\hat H}\Phi\big]
(1-\varpi_{F;\hat H}),
&\cr
&\vdots.
&(2.42)\cr}
$$
As functional integration path, it is convenient to use
$(g_t,G_t)=(\exp(t\phi)\hat h,\exp(t\Phi)\hat H)$ with $0\leq t\leq 1$.
One then finds
$$
\eqalignno{
&S_{w,F}(h,H;\hat h,\hat H)
=-{(6w^2-6w+1)r_F\over 6\pi}\int_\Sigma d^2z
\Big[{1\over 2}\bar\partial\phi\partial\phi-f_{\hat h}\phi\Big]
&\cr
&-{(2w-1)\over 2\pi}\int_\Sigma d^2z\Big[
{1\over 2}\bar\partial\phi\partial\tr D_{F;\hat H}(\Phi)
+{1\over 2}\partial\phi\bar\partial\tr D_{F;\hat H}(\Phi)
-f_{\hat h}\tr D_{F;\hat H}(\Phi)
&\cr
&\hskip6cm
-\tr\big((F_{\hat H}+\bar\partial\partial_{\hat H}\varpi_{F;\hat H})
\varpi_{F;\hat H}\big)\phi\Big]
&\cr
&-{1\over\pi}\int_\Sigma d^2z\tr\Big[
\partial_{\hat H}\Phi K^*_{F;\hat H}(\Phi,\bar\partial\Phi)
-F_{\hat H}D_{F;\hat H}(\Phi)+T_{F;\hat H}(\Phi)\Big],
&(2.43)\cr}
$$
where
$$\eqalignno{
D_{F;\hat H}(\Phi)
=&\sum_{m=0}^\infty{1\over(m+1)!}\sum_{n=0}^m{m\choose n}
\varpi^{(m-n)}_{F;\hat H}(\Phi)\Phi\varpi^{(n)}_{F;\hat H}(\Phi),
&(2.44)\cr
T_{F;\hat H}(\Phi)
=&\sum_{m=0}^\infty{1\over(m+1)!}\sum_{n=0}^m{m\choose n}
\partial_{\hat H}\varpi^{(m-n)}_{F;\hat H}(\Phi)\Phi
\bar\partial\varpi^{(n)}_{F;\hat H}(\Phi),
&(2.45)\cr
K^*_{F;\hat H}(\Phi,\bar\partial\Phi)
=&\sum_{m=0}^\infty{1\over(m+2)!}\sum_{n=0}^m{m+1\choose n}
(-\ad\Phi)^{m-n}\sum_{k=0}^n{n\choose k}\bar\partial
\big(\varpi^{(n-k)}_{F;\hat H}(\Phi)\Phi\big)\varpi^{(k)}_{F;\hat H}(\Phi).
&\cr
&&(2.46)\cr}
$$
The above expression greatly simplifies when $F=E$, since in such case
$\varpi^{(r)}_{F;\hat H}(\Phi)=\delta_{r,0}1$. One then finds
$$
\eqalignno{
&S_{w,F}(h,H;\hat h,\hat H)
=-{(6w^2-6w+1)r_E\over 6\pi}\int_\Sigma d^2z
\Big[{1\over 2}\bar\partial\phi\partial\phi-f_{\hat h}\phi\Big]
&\cr
&-{(2w-1)\over 2\pi}\int_\Sigma d^2z\Big[
{1\over 2}\bar\partial\phi\partial\tr\Phi
+{1\over 2}\partial\phi\bar\partial\tr\Phi
-f_{\hat h}\tr\Phi-\tr(F_{\hat H})\phi\Big]
&\cr
&-{1\over\pi}\int_\Sigma d^2z\tr\Big[
\bar\partial\Phi{\exp\ad\Phi-1-\ad\Phi\over(\ad\Phi)^2}\partial_{\hat H}\Phi
-F_{\hat H}\Phi\Big].
&(2.47)\cr}
$$

Expression $(2.43)$ provides the appropriate generalization of the Liouville
action in the present context. By setting $H=\hat H$ and $\Phi=0$ in
$(2.43)$, one recovers in fact the customary conformal anomaly. The central
charge is
$$
c_{w,F}=-2(6w^2-6w+1)r_F
\eqno(2.48)
$$
and is the same as that of $r_F$ copies of a spin $w$ fermionic $b-c$
system. By setting $h=\hat h$ and $\phi=0$, one gets the generalized
Weyl anomaly. When $F=E$, so that $(2.47)$ holds, such anomaly is given
by the Donaldson action as discussed in ref.\ref{12}.

To conclude this section, let us discuss renormalization.
{}From $(2.35)$, it appears that in order to renormalize
$I^{\rm bare}_{w,F}(h,H;\epsilon)$, one has to add a counterterm of the
form
$$
\Delta I^{\rm bare}_{w,F}(h,H;\epsilon)
=\lambda_{\rm bare}(\epsilon)\int_\Sigma d^2zh+\nu_{\rm bare}(\epsilon)
+\Delta I^{\rm ren}_{w,F}(h,H)+O(\epsilon).
\eqno(2.49)
$$
Here,
$$
\eqalignno{
\lambda_{\rm bare}(\epsilon)=&~{r_F\over\pi\epsilon}+\lambda_{\rm ren}
+O(\epsilon),
&(2.50)\cr
\nu_{\rm bare}(\epsilon)
=&-K_{w,F}\ln\epsilon+\nu_{\rm ren}+O(\epsilon).
&(2.51)\cr}
$$
$\Delta I^{\rm ren}_{w,F}(h,H)$ is a finite local functional of $(h,H)$.
Its choice defines a renormalization prescription. The renormalized
effective action is
$$
I^{\rm ren}_{w,F}(h,H)
=\lim_{\epsilon\rightarrow 0}\Big[
I^{\rm bare}_{w,F}(h,H;\epsilon)+\Delta I^{\rm bare}_{w,F}(h,H;\epsilon)
\Big].
\eqno(2.52)
$$
{}From $(2.35)$ and $(2.49)$, one has
$$
I^{\rm ren}_{w,F}(h,H)
=\lambda_{\rm ren}\int_\Sigma d^2zh+\nu_{\rm ren}
+S_{w,F}(h,H;\hat h,\hat H)+s_{w,F}(\hat h,\hat H)
+\Delta I^{\rm ren}_{w,F}(h,H).
\eqno(2.53)
$$
{}From $(2.53)$ and $(2.36)$, one obtains then
$$
\delta I^{\rm ren}_{w,F}(h,H)=
\lambda_{\rm ren}\int_\Sigma d^2z\delta h
+A_{w,F}(h,H),
\eqno(2.54)
$$
where
$$
A_{w,F}(h,H)=A^0_{w,F}(h,H)+\delta\Delta I^{\rm ren}_{w,F}(h,H)
\eqno(2.55)
$$
with $A^0_{w,F}(h,H)$ given by $(2.34)$. $(2.54)-(2.55)$ provide the
expression of the generalized Weyl anomaly. Note that the anomaly is local
since $\varpi_{F;H}$ is a local functional of $H$ by $(2.19)$.

If minimal subtraction is applied, one has $\lambda_{\rm ren}=\nu_{\rm ren}
=0$ and $\Delta I^{\rm ren}_{w,F}(h,H)=0$. Another possibility is to have
$\lambda_{\rm ren}\not=0$. This would lead to a generalization of the
Liouville model. Other interesting
renormalizations may be considered in specific models,
such as the DS ghost system discussed in detail in sect. 4.

Extended conformal field theory, studied in
ref. \ref{12}, is a particular case of the above framework with $F=E$ and
$w={1\over 2}$. The case where $F=E$ but $w\not={1\over 2}$, can be reduced
to the latter one by redefining the bundles
$E$ and $F$ into $E_w=k^{\otimes w-{1\over 2}}\otimes E$
and $F_w=k^{\otimes w-{1\over 2}}\otimes F$, respectively, and $w$ into
$1\over 2$ and the hermitian structure $(h,H)$ into
$(h,h^{\otimes w-{1\over 2}}\otimes H)$.
\vskip.4cm
\item{3.} $F${\bf--special holomorphic structures and} $F${\bf--special
automorphisms}
\vskip.4cm
\par
Let $E$ be a smooth vector bundle of rank $r_E$ over a compact smooth
surface of genus $\ell$. Let further $F$ be a subbundle of $E$
of rank $r_F>0$.

Let $\goth S_E$ be the family of holomorphic structures $\sans S$ of $E$.
For $\sans S\in\goth S_E$, let $E_{\ssans S}$ be the corresponding holomorphic
vector bundle. In general, there does not exist a subbundle $F_{\ssans S}$
of $E_{\ssans S}$
corresponding to $F$. The holomorphic structures $\sans S\in\goth S_E$,
for which this happens, are called $F$--special. They form a subfamily
$\goth S_F$ of $\goth S_E$. The holomorphic structures $\sans S\in\goth S_F$
are precisely those, for which the formalism developed in sect. 2 applies.

Let $\Aut_{cE}$ and $\Diff_c$ be be the groups of smooth automorphisms of $E$
homotopic to $\id_E$ and of smooth diffeomorphisms of $\Sigma$ homotopic to
$\id_\Sigma$, respectively. If $\alpha\in\Aut_{cE}$, there exists
$f_\alpha\in\Diff_c$ such that $\pi\circ\alpha=
f_\alpha\circ\pi$ and that $\alpha|_{E_p}$ is a linear isomorphisms
of $E_p$ onto $E_{f_\alpha(p)}$ for every $p\in\Sigma$, where $\pi$ is
the bundle projection and $E_p$ is the fiber of $E$ at $p$. In general,
for a given $\alpha\in\Aut_{cE}$, $\alpha|_{E_p}$ does not map $F_p$ onto
$F_{f_\alpha(p)}$, {\it i. e.} $\alpha$ does not respect the subbundle
$F$. The automorphisms $\alpha\in\Aut_{cE}$, for which this happens, are
called $F$--special. They form a subgroup $\Aut_{cF}$ of $\Aut_{cE}$.
This is the relevant symmetry group for the field theoretic
constructions of sect. 2.

There is a natural action of $\Aut_{cE}$ on $\goth S_E$. This associates to
any $\sans S\in\goth S_E$ and any $\alpha\in\Aut_{cE}$ the pull back
$\alpha^*\sans S\in\goth S_E$
of $\sans S$ by $\alpha$ (see \ref{12} for a detailed discussion).
A simple but important theorem states that $\Aut_{cF}$ preserves
$\goth S_F$.
\par\noindent
{\it Proof}. Let $\sans S\in\goth S_E$. $\sans S$ is a collection of
trivializations
$\{(z_{\ssans S a},u_{\ssans S a})\}$, where $z_{\ssans S a}$ is a
complex coordinate on $\Sigma$, $u_{\ssans S a}$ is a fiber coordinate and,
whenever defined, $\bar\partial_{\ssans S a}z_{\ssans S b}=0$ and
$u_{\ssans S a}=T_{\ssans S ab}\circ\pi u_{\ssans S b}$ with
$T_{\ssans S ab}$ an $r_E\times r_E$ matrix valued function such that
$\bar\partial_{\ssans S c}T_{\ssans S ab}=0$.
If $\alpha\in\Aut_{cE}$ and $\sans S\in\goth S_E$,
then $\alpha^*\sans S\in\goth S_E$ with $z_{\alpha^*\ssans S a}=
z_{\ssans S b}\circ f_\alpha$ and $u_{\alpha^*\ssans S a}=
u_{\ssans S b}\circ\alpha$ and $T_{\alpha^*\ssans S ac}=T_{\ssans S bd}
\circ f_\alpha$ for suitably related $a$, $c$ and $b$, $d$.
If $\sans S\in\goth S_F$,
there exists, for each trivialization $(z_{\ssans S a},u_{\ssans S a})$,
an $r_E\times r_E$ matrix valued function $\Theta_{\ssans S a}$ such that
$(\Theta_{\ssans S a}\circ\pi u_{\ssans S a})(F\cap\pi^{-1}
(\dom z_{\ssans S a}))$ has the last $r_E-r_F$ components identically zero,
$\Theta_{\ssans S a}T_{\ssans S ab}\Theta_{\ssans S b}{}^{-1}$ has vanishing
lower left $(r_E-r_F)\times r_F$ block and $\bar\partial_{\ssans S a}
\Theta_{\ssans S a}=0$.
If $\alpha\in\Aut_{cF}$ and $\sans S\in\goth S_F$, then
it is straightforward to verify that $\alpha^*\sans S\in\goth S_F$ by
setting $\Theta_{\alpha^*\ssans S a}=\Theta_{\ssans S b}\circ f_\alpha$.
\hfill{\it QED}

Let $\sans S\in\goth S_E$ be a holomorphic structure of $E$ and
$(h_{\ssans S},H_{\ssans S})$ be a hermitian structure on $E_{\ssans S}$.
If $\alpha\in\Aut_{cE}$, then the pull--back
$(\alpha^*h_{\alpha^*\ssans S},\alpha^*H_{\alpha^*\ssans S})$ of
$(h_{\ssans S},H_{\ssans S})$ by $\alpha$ is a hermitian structure on
$E_{\alpha^*\ssans S}$.

Let ${\cal H}_{w,\bar w,F;h,H;\ssans S}$ be the Hilbert space defined
in sect. 1 with the holomorphic structure $\sans S\in\goth S_F$ indicated.
If $\alpha\in\Aut_{cF}$, then the pull-back operator $\alpha^*$ is
a unitary operator of ${\cal H}_{w,\bar w,F;h,H;\ssans S}$ onto
${\cal H}_{w,\bar w,F;\alpha^*h,\alpha^*H;\alpha^*\ssans S}$
\par\noindent
{\it Proof}. This follows easily from $(2.1)$ using the relations
$\alpha^*\phi_{a\alpha^*\ssans S}=\phi_{b\ssans S}\circ f_\alpha$ with
$\phi_{\ssans S}\in{\cal H}_{w,\bar w,F;h,H;\ssans S}$ and
$\alpha^*h_{a\alpha^*\ssans S}=(h_{b\ssans S}\circ f_\alpha$ and
$\alpha^*H_{a\alpha^*\ssans S}=H_{b\ssans S}\circ f_\alpha)$
for suitably related $a$ and $b$. \hfill{\it QED}

If $\sans S\in\goth S_F$ and $\alpha\in\Aut_{cF}$, then
$\bar\partial_{w,F;\alpha^*\ssans S}
=\alpha^*\circ\bar\partial_{w,F;\ssans S}\circ\alpha^{*-1}$. This implies,
among other things, that
$\Delta^\sharp_{w,F;\alpha^*h,\alpha^*H;\alpha^*\ssans S}
=\alpha^*\circ\Delta^\sharp_{w,F;h,H;\ssans S}\circ\alpha^{*-1}$.
$\alpha^*$ being unitary, the spectrum of $\Delta^\sharp_{w,F;h,H;\ssans S}$
is $\Aut_{cF}$ invariant.

The above geometrical treatment is elegant but abstract. One would like to
translate it into the language of field theory, which is the one suitable
for physical applications. This can be achieved as follows \ref{12}.

For any pair of holomorphic structures $\sans S_1$, $\sans S_2$, there
exist two distinguished sections $\lambda_{\ssans S_1\ssans S_2}$ and
$V_{\ssans S_1\ssans S_2}$ of
$k_{\ssans S_1}\otimes k_{\ssans S_1}{}^{\otimes-1}$ and
$E_{\ssans S_1}\otimes E_{\ssans S_2}{}^\vee$, respectively,
called intertwiners.
Write a generic trivialization of $\sans S_i$ as
$(z_{\ssans S_i},u_{\ssans S_i})$, where $z_{\ssans S_i}$ is a
complex coordinate on $\Sigma$ and $u_{\ssans S_i}$ is a fiber
coordinate. Then, $\lambda_{\ssans S_1\ssans S_2}=\partial_{\ssans S_1}
z_{\ssans S_2}$ and $V_{\ssans S_1\ssans S_2}$ is defined by the relation
$u_{\ssans S_1}=V_{\ssans S_1\ssans S_2}\circ\pi u_{\ssans S_2}$.
The intertwiners define an isomorphism between the space of sections of
each vector bundle constructed by means of $k_{\ssans S_1}$ and
$E_{\ssans S_1}$ and the space of sections of the corresponding bundle
constructed by means of $k_{\ssans S_2}$ and $E_{\ssans S_2}$.
Hence, the field content of a field theory having $E$ as topological
background is described completely by the spaces of sections of
vector bundles built by means of $k_{\ssans S_0}$ and
$E_{\ssans S_0}$ for a fiducial reference holomorphic structure
$\sans S_0$.
All relevant field theoretic relations may be thus written
in terms of the trivializations of $\sans S_0$.
By convention, when a field or a combination of
fields carries no subscript $\sans S$, then it is represented in
terms of ${\sans S}_0$. Note that by $E$ and $k$ it is denoted both
the holomorphic vector bundle $E_{\ssans S_0}$ and canonical line bundle
$k_{\ssans S_0}$ and their smooth counterparts.
This generates no confusion since from the context
it will be clear which is meant.

There is a one--to--one correspondence between the family $\goth S_E$ of
holomorphic structures $\sans S$ of $E$ and the family of pairs $(\mu,A^*_A)$,
where $\mu$ is a Beltrami field and $A^*_A$ is a Koszul field \ref{12}. Recall
that a Beltrami field $\mu$ is an element of ${\cal S}_{-1,1}$ such that
$\sup_\Sigma|\mu|<1$ and that a Koszul field $A^*_A$ is an element of
${\cal S}_{0,1,\End E}$. For $\sans S=(\mu,A^*_A)\in\goth S_E$, one has
$$
\mu=\bar\partial z_{\ssans S}/\partial z_{\ssans S},
\eqno(3.1)
$$
$$
A^*_A=(\bar\partial-\mu\partial)V_{\ssans S}V_{\ssans S}{}^{-1}+\mu A,
\eqno(3.2)
$$
where $A$ is a fixed $(1,0)$ connection of $E$, $\bar\partial\equiv
\bar\partial_{\ssans S_0}$ and $V_{\ssans S}\equiv V_{\ssans S_0\ssans S}$
\ref{12}.

All the identities of sect. 1, valid for an arbitrary holomorphic
structure $\sans S\in\goth S_F$, may be easily written in the
Beltrami--Koszul parametrization by performing the formal substitutions
$$
d^2z\rightarrow d^2z(1-\bar\mu\mu),
\eqno(3.3)
$$
$$
\bar\partial\rightarrow
{1\over (1-\bar\mu\mu)}(\bar\partial-\mu\partial-w\partial\mu),
\quad{\rm on~} {\cal S}_{w,0},
\eqno(3.4)
$$
$$
\bar\partial\rightarrow
{1\over (1-\bar\mu\mu)}
\big(\bar\partial-\mu\partial_H-w\partial\mu-\ad A^*_H\big),
\quad{\rm on~} {\cal S}_{w,0,\End E},
\eqno(3.5)
$$
$$
f_h\rightarrow
{1\over (1-\bar\mu\mu)}\bigg[
f_h-\big(\partial-\bar\mu\bar\partial-\bar\partial\bar\mu\big)
{\partial_h\mu\over 1-\bar\mu\mu}
-\big(\bar\partial-\mu\partial-\partial\mu\big)
{\bar\partial_h\bar\mu\over 1-\bar\mu\mu}
\bigg],
\eqno(3.6)
$$
$$
\eqalignno{
F_H\rightarrow
{1\over (1-\bar\mu\mu)}\bigg[
F_H-\big(\partial_H-\bar\mu\bar\partial-\bar\partial\bar\mu\big)
{A^*_H\over 1-\bar\mu\mu}
-\big(\bar\partial-
&\mu\partial_H-\partial\mu\big)
{HA^*_H{}^\dagger H^{-1}\over 1-\bar\mu\mu}
&\cr
&+{[A^*_H,HA^*_H{}^\dagger H^{-1}]\over 1-\bar\mu\mu}\bigg].
&(3.7)\cr}
$$
Here, $\partial_h=\partial+\partial\ln h$ is the covariant derivative
associated to the metric connection $\partial\ln h$ acting on
${\cal S}_{-1,1}$.
$A^*_H$ is given by $(3.2)$ with $A=\partial HH^{-1}$.
By using the Beltrami--Koszul parametrization one may also check that the
the integral expression $(2.32)$ is independent from the holomorphic
structure chosen, as expected from the index theorem.

The Beltrami--Koszul parametrization allows one to state a condition for
a holomorphic structure $\sans S\in\goth S_E$ to be $F$--special.
$\sans S\in\goth S_F$ if and only if
$$
\big(\bar\partial\varpi_{F;H}\varpi_{F;H}\big)_{\ssans S}=0.
\eqno(3.8)
$$
\par\noindent {\it Proof}.
This condition is necessary, as explained in sect. 2. It is also sufficient.
For if $(3.8)$ holds, there exists on each trivialization domain a local
holomorphic frame in $E_{\ssans S}$ spanning $F$, implying that
$E_{\ssans S}$ contains a holomorphic subbundle $F_{\ssans S}$ corresponding
to $F$. \hfill{\it QED}
\par\noindent
The dependence of
this condition on the metric $H$ is only apparent. In fact, using $(2.19)$
it is easy to show that if $\sans S$ satisfies $(3.8)$ for a given hermitian
metric $H$, then it does also satisfy it for any other close metric $H'$.
It must clearly be so, for the space $\goth S_F$, by its definition, does not
depend on a choice of hermitian structure. $(3.8)$ can be written
in the Beltrami--Koszul parametrization, where it reads
$$
\big(\bar\partial-\mu\partial_A-\ad A^*_A\big)\varpi_{F;H}\varpi_{F;H}=0,
\eqno(3.9)
$$
where $\partial_A=\partial-\ad A$ is the covariant derivative on $\End E$
associated to $A$.
This is the field theoretic constraint that must be obeyed by a holomorphic
structure $\sans S=(\mu,A^*_A)$ in order it to belong to $\goth S_F$.

In the analysis of symmetries, it is much simpler to proceed at the
infinitesimal level. Let $s$ be the nilpotent Slavnov operator, $s^2=0$.
Let $c$ and $m$ be the automorphisms ghosts \ref{12}. $c$ is the
diffeomorphism ghost associated to the natural map $\Aut_{cE}\rightarrow
\Diff_c$ defined earlier. $c$ is a section of $k^{-1}$ valued in
$\bigwedge^1(\Lie\Aut_{cE})^\vee$. $M$ corresponds to the action of
$\Aut_{cE}$ on the fibers of $E$. For a given background $(1,0)$ connection
of $E$, $M-cA$ is a section of $\End E$ valued in
$\bigwedge^1(\Lie\Aut_{cE})^\vee$. The Maurer--Cartan equations of
$\Aut_{cE}$ yield
$$
sc=(c\partial+\bar c\bar\partial)c,
\eqno(3.10)
$$
$$
sM=(c\partial+\bar c\bar\partial)M-{1\over 2}[M,M]
\eqno(3.11)
$$
\ref{12}. The action of $\Aut_{cE}$ on $\goth S_E$ induces an action on the
Beltrami--Koszul fields $(\mu,A^*_A)$ given by
$$
s\mu=\big(\bar\partial-\mu\partial+\partial\mu\big)C,
\eqno(3.12)
$$
$$
sA^*_A=\big(\bar\partial-\mu\partial_A-\ad A^*_A\big)X_A
+C\big(\partial_AA^*_A-\bar\partial A\big),
\eqno(3.13)
$$
where
$$
C=c+\mu\bar c,
\eqno(3.14)
$$
$$
X_A=cA+\bar cA^*_A-M.
\eqno(3.15)
$$
$C$ and $X_A$ are sections of $k^{-1}$ and $\End E$ valued in
$\bigwedge^1(\Lie\Aut_{cE})^\vee$ and depending on $(\mu,A^*_A)$,
respectively. Further
$$
sC=C\partial C,
\eqno(3.16)
$$
$$
sX_A=C\partial_AX_A+{1\over 2}[X_A,X_A].
\eqno(3.17)
$$

The pull-back action of $\Aut_{cE}$ on the space $\goth H_E$ of hermitian
structures $(h,H)$ of $E$ yields
$$
s\ln h=(c\partial+\bar c\bar\partial)\ln h
+\partial c+\mu\partial\bar c+\bar\partial\bar c+\bar\mu\bar\partial c,
\eqno(3.18)
$$
$$
sHH^{-1}=(c\partial+\bar c\bar\partial)HH^{-1}-M-HM^\dagger H^{-1}.
\eqno(3.19)
$$

The action of the $F$--special automorphism group $\Aut_{cF}$ can still
be expressed at the infinitesimal level by means of the automorphism ghost
fields $c$ and $M$. The restriction to $\Lie\Aut_{cF}$ shows up as a
relation obeyed by $c$ and $M$, which will be derived in a moment.
It is not difficult to show that an automorphism
$\alpha\in\Aut_{cE}$ belongs to $\Aut_{cF}$ if and only if
$$
\varpi_{F;\alpha^* H}=\alpha^*\varpi_{F;H}.
\eqno(3.20)
$$
\par\noindent{\it Proof}.
Denote by $(z_a,u_a)$ the trivializations of the reference holomorphic
structure, as done earlier. If $\alpha\in\Aut_{cE}$, then, for any two
trivializations $(z_a,u_a)$ and $(z_b,u_b)$ such that $\dom z_a\cap
f_\alpha(\dom z_b)\not=\emptyset$, there exists a local $r_E\times r_E$
smooth matrix function $\hat\alpha_{ab}$ such that $u_a\circ\alpha=
\hat\alpha_{ab}\circ\pi u_b$. One further has $\alpha^*\Theta_b=
\hat\alpha_{ab}{}^{-1}\Theta_a\circ f_\alpha\hat\alpha_{ab}$ for any
element $\Theta$ of ${\cal S}_{0,0,\End E}$. Let $\alpha\in\Aut_{cF}$.
Then, for any $x\in F$, one has $(\varpi_{F;H}\circ\pi u)(\alpha(x))=
u(\alpha(x))$, since $\alpha(x)\in F$.
This implies that, for any $x\in F$,
$(\alpha^*\varpi_{F;H}\circ\pi u)(x)=u(x)$. Then, since $\alpha^*\varpi_{F;H}$
is $\alpha^*H$--hermitian, $(3.20)$ holds. Next, let $\alpha\in\Aut_{cE}$
satisfy $(3.20)$. Then, since $(\varpi_{F;\alpha^*H}\circ\pi u)(x)=u(x)$ for
any $x\in F$, one has that $(\alpha^*\varpi_{F;H}\circ\pi u)(x)=u(x)$
for $x\in F$. This implies that for any $x\in F$, one has
$(\varpi_{F;H}\circ\pi u)(\alpha(x))=u(\alpha(x))$. Hence, for any $x\in F$,
$\alpha(x)\in F$, so that $\alpha\in\Aut_{cF}$.
\hfill{\it QED}
\par\noindent
Using $(2.19)$, one can also show that this condition is actually
independent from the metric $H$, as expected on general grounds
form the metric independence of $\Aut_{cF}$.
Going over the infinitesimal formulation and using $(2.19)$ and $(3.19)$,
one finds that, for the $F$--special symmetry,
$$
\big(c\partial+\bar c\bar\partial-\ad M\big)\varpi_{F;H}\varpi_{F;H}=0,
\eqno(3.21)
$$
which is the constraint on $c$ and $M$ looked for.
If $(\mu,A^*_A)$ is an $F$--special holomorphic structure, so that $(3.9)$
is fulfilled, then $(3.21)$ can be stated in terms of the
$(\mu,A^*_A)$ dependent ghost fields
$C$ and $X_A$ given by $(3.14)$ and $(3.15)$ as follows
$$
\big(C\partial_A+\ad X_A)\varpi_{F;H}\varpi_{F;H}=0.
\eqno(3.22)
$$

It can be verified that $(3.21)$ is compatible with $(3.10)$ and $(3.11)$
in the following sense. If one applies $s$ to the left hand side of
$(3.21)$ and uses $(3.10)$, $(3.11)$ and $(2.19)$, one obtains a result that
is linear in the left hand side of $(3.21)$. Thus, enforcing the constraint
$(3.21)$ is compatible with the action $\Aut_{cE}$. This is expected on general
grounds and verified here. Similarly, if one applies $s$ to the left hand side
of $(3.9)$ and uses $(3.12)$, $(3.13)$ and $(2.19)$, one obtains an expression
linear in the left hand sides of $(3.9)$ and $(3.22)$. Hence, imposing the
constraints $(3.9)$ and $(3.22)$ is again compatible with the action of
$\Aut_{cE}$.

In the Beltrami--Koszul parametrization of holomorphic structures the
determinant of $\Delta^\sharp_{w,F;h,H;\ssans S}$ and the associated
bare and renormalized effective actions become functionals of the geometrical
fields $\mu$, $\bar\mu$, $A^*_H$ and $A^*_H{}^\dagger$. The $\Aut_{cF}$
invariance of the spectrum of $\Delta^\sharp_{w,F;h,H;\ssans S}$ implies
that its determinant also is invariant. Hence
$$
s\dtr'_\epsilon\Delta^\sharp_{w,F;h,H;\mu,\bar\mu,A^*_H,A^*_H{}^\dagger}=0.
\eqno(3.23)
$$
The bare effective action $I^{\rm bare}_{w,F}(h,H;\sans S;\epsilon)$ cannot
really be considered a functional over the space $\goth H_E\times\goth S_F$
because of the ambiguity inherent in the choice of the bases of zero modes.
For this reason, $I^{\rm bare}_{w,F}(h,H;\sans S;\epsilon)$
is invariant under $\Aut_{cF}$ only up to redefinitions of
the zero mode bases. However, the exponential of $I^{\rm bare}_{w,F}(h,H;
\sans S;\epsilon)$ can be viewed as a section of a line bundle on
$\goth H_E\times\goth S_F$. As such,
$I^{\rm bare}_{w,F}(h,H;\sans S;\epsilon)$ is in fact $\Aut_{cF}$ invariant
and one has
$$
sI^{\rm bare}_{w,F}(h,H;\mu,\bar\mu,A^*_H,A^*_H{}^\dagger;\epsilon)=0.
\eqno(3.24)
$$
The exponential of the renormalized effective action
$I^{\rm ren}_{w,F}(h,H;\sans S)$ may be viewed
similarly as a section of the same
line bundle on $\goth H_E\times\goth S_F$.
The counterterm $\Delta I^{\rm bare}_{w,F}(h,H;\sans S;\epsilon)$ given by
$(2,49)$ is $\Aut_{cF}$ invariant if $\Delta I^{\rm ren}_{w,F}(h,H;
\sans S)$ is. In that case case, $I^{\rm ren}_{w,F}(h,H;\sans S)$, also,
is $\Aut_{cF}$ invariant and one has
$$
sI^{\rm ren}_{w,F}(h,H;\mu,\bar\mu,A^*_H,A^*_H{}^\dagger)=0.
\eqno(3.25)
$$
\vskip.4cm
\item{4.} {\bf The Drinfeld--Sokolov ghost system}
\vskip.4cm
\par
The basic algebraic data entering in the definition
of the model are the following: {\it i}) a simple complex Lie group $G$;
{\it ii}) an $SL(2,\Bbb C)$ subgroup $S$ of $G$ invariant under the
compact conjugation $\dagger$ of $G$.
Let $t_{-1}$, $t_0$, $t_{+1}$ be a set of standard generators
of $\goth s$, {\it i. e.}
$$
[t_{+1},t_{-1}]=2t_0,\quad [t_0,t_{\pm 1}]=\pm t_{\pm 1},
\eqno(4.1)
$$
$$
t_d{}^\dagger=t_{-d}, \quad d=-1,0,+1.
\eqno(4.2)
$$
To the Cartan element $t_0$ of $\goth s$, there is associated a
halfinteger grading of $\goth g$: the subspace $\goth g_m$ of $\goth g$
of degree $m\in \Bbb Z/2$ is the eigenspace of $\ad t_0$ with
eigenvalue $m$. One can further define a
bilinear form $\chi$ on $\goth g$ by $\chi(x,y)=\tr_{\rm ad}(t_{+1}[x,y])$,
$x,y\in \goth g$ \ref{14}, where $\tr_{\rm ad}$ denotes the Cartan--Killing
form. The restriction of $\chi$ to
$\goth g_{-{1\over 2}}$ is non singular. By Darboux theorem,
there is a direct sum decomposition $\goth g_{-{1\over 2}}=
\goth p_{-{1\over 2}}\oplus\goth q_{-{1\over 2}}$ of $\goth g_{-{1\over 2}}$
into subspaces of the same dimension, which are maximally isotropic and dual
to each other with respect to $\chi$. Set
$$
\goth x=\goth p_{-{1\over 2}}\oplus\bigoplus_{m\leq -1}\goth g_m.
\eqno(4.3)
$$
$\goth x$ is a negative graded nilpotent subalgebra of $\goth g$.

On a Riemann surface $\Sigma$ of genus $\ell$ with a spinor
structure $k^{\otimes {1\over 2}}$, one may define the $G$ valued
holomorphic $1$--cocycle
$$
L_{ab}=\exp(-\ln k_{ab}t_0)\exp(\partial_a k_{ab}{}^{-1}t_{-1}).
\eqno(4.4)
$$
This in turn defines a holomorphic principal $G$--bundle, the
Drinfeld--Sokolov (DS) bundle \ref{16,24}.
$\Ad L$ is one of the associated holomorphic vector bundles.
The $\goth x$--valued sections of $\Ad L$ span a subbundle
$\Ad L_{\goth x}$ of $\Ad L$, since $\goth x$ is invariant under
$\ad t_0$ and $\ad t_{-1}$.

The DS ghost system $\beta-\gamma$, described in the introduction, is governed
by the action $(1.1)$, where $\beta$ and $\gamma$ are anticommuting sections
of $k\otimes\Ad L$ and $\Ad L$ valued in $\goth g/\goth x^\perp$ and
$\goth x$, respectively, $\goth x^\perp$ being the orthogonal complement of
$\goth x$ with respect to $\tr_{\rm ad}$.
The effective action of the DS ghost system is thus of the type described
in sect. 2 with $E=\Ad L$, $F=\Ad L_{\goth x}$ and $w=0$. The hermitian
structures of $\Ad L$ considered here are of the form $(h,\Ad H)$, where
$H$ is a hermitian metric of $L$. From $(2.54)$, $(2.55)$ and $(2.34)$,
one finds that the renormalized effective action
$I^{\rm ren}_{DS}(h,H)$ satisfies the Weyl anomalous Ward identity
$$
\delta I^{\rm ren}_{DS}(h,H)=\lambda_{\rm ren}\int_\Sigma d^2z\delta h+
A_{DS}(h,H),
\eqno(4.5)
$$
$$
A_{DS}(h,H)=A^0_{DS}(h,H)+\delta\Delta I^{\rm ren}_{DS}(h,H),
\eqno(4.6)
$$
where
$$
\eqalignno{
&A^0_{DS}(h,H)={r_{DS}\over 6\pi}\int_\Sigma d^2z\delta\ln hf_h
&\cr
&-{1\over 2\pi}\int_\Sigma d^2z\Big[\delta\ln h
\tr\big((\ad F_H+\bar\partial\partial_H\varpi_H)\varpi_H\big)
+\tr\big(\ad(\delta HH^{-1})\varpi_H\big)f_h\Big]&\cr
&+{1\over\pi}\int_\Sigma d^2z
\tr\big(\ad(\delta HH^{-1})(\ad F_H+\bar\partial\partial_H\varpi_H)
\varpi_H\big),
&(4.7)\cr}
$$
with $r_{DS}=\dim\goth x$
and $\Delta I^{\rm ren}_{DS}(h,H)$ is a local functional of $(h,H)$
\footnote{${}^2$}{In this section, I shall suppress the indices $w$ and $F$
to lighten the notation.}.

For the DS principal bundle $L$, there exists a distinguished choice
of the fiber metric $H$ for any given hermitian metric $h$ on the base
$\Sigma$, namely
$$
H_h=\exp(-\partial\ln ht_{-1})\exp(-\ln h t_0)\exp(-\bar\partial\ln ht_{+1}).
\eqno(4.8)
$$
It is not difficult to show that the corresponding projector $\varpi_{H_h}$
is given by
$$
\varpi_{H_h}
=\exp(-\partial\ln h\ad t_{-1})p_{\goth x}\exp(\partial\ln h\ad t_{-1}),
\eqno(4.9)
$$
where $p_{\goth x}$ is the orthogonal projector of $\goth g$ onto $\goth x$
with respect to the hermitian inner product on $\goth g$ defined by
$(x,y)=\tr_{\rm ad}(x^\dagger y)$ for $x,y\in\goth g$.

For the DS ghost system, besides the minimal subtraction
renormalization prescription, corresponding to setting
$\Delta I^{\rm ren}_{DS}(h,H)=0$, there is another relevant renormalization
defined by the choice
$$
\Delta I^{\rm ren}_{DS}(h,H)
={1\over 2\pi}\int_\Sigma d^2z\Big[\int_{H_h}^H\tr\big(
\ad(\delta H'H'^{-1})\varpi_{H'}\big)\Big]f_h.
\eqno(4.10)
$$
The functional $1$--form $\tr\big(\ad(\delta HH^{-1})\varpi_H\big)$
of $\Omega^1_{\goth H_L}$ is exact. Thus, the above functional line integral
does not depend on the choice of the functional path joining
$H_h$ to $H$ in $\goth H_L$.
Using the Taylor expansion $(2.41)$ and $(4.8)$ and $(4.9)$, one can verify
that $\Delta I^{\rm ren}_{DS}(h,H)$ is a local functional of $(h,H)$.
Further, using $(2.31)$, one can show that
$$
\eqalignno{
&\delta\Delta I^{\rm ren}_{DS}(h,H)
={1\over 2\pi}\int_\Sigma d^2z\Big[\delta\ln h
\tr\big((\ad F_H+\bar\partial\partial_H\varpi_H)\varpi_H\big)
+\tr\big(\ad(\delta HH^{-1})\varpi_H\big)f_h\Big]
&\cr
&-{1\over 2\pi}\int_\Sigma d^2z\Big[\delta\ln h
\tr\big((\ad F_{H_h}+\bar\partial\partial_{H_h}\varpi_{H_h})\varpi_{H_h}\big)
+\tr\big(\ad(\delta H_hH_h{}^{-1})\varpi_{H_h}\big)f_h\Big].
&(4.11)\cr}
$$
Hence, on account of $(4.5)-(4.7)$,
choosing $\Delta I^{\rm ren}_{DS}(h,H)$ to be given by $(4.10)$,
one obtains a renormalized effective action $I^{\rm ren}_{DS}(h,H)$, for which
the classical $H$ equations are of the form
$$
\tr\big(\ad(\delta HH^{-1})(\ad F_H+\bar\partial\partial_H\varpi_H)
\varpi_H\big)+\ldots=0.
\eqno(4.12)
$$
The ellipses denote terms coming from the matter sector of the model,
which will not be discussed here \ref{15}. The relevant point is that
the above classical equations, including the contributions coming from
the matter sector not shown, do not contain the surface metric $h$.
Thus, the classical $H$ dynamics induced by $I^{\rm ren}_{DS}(h,H)$ is
conformally invariant.

With the metric $H_h$ available, one may define the reduced
renormalized effective action
$$
I^{\rm ren}_{DS}(h)=I^{\rm ren}_{DS}(h,H_h)
\eqno(4.13)
$$
for any choice of the renormalization prescription. Here,
$\Delta I^{\rm ren}_{DS}(h,H)$ is meaningfully chosen to be of the form
$$
\Delta I^{\rm ren}_{DS}(h,H)
={\kappa_0\over\pi}\int_\Sigma d^2z h^{-1}f_h{}^2,
\eqno(4.14)
$$
where $\kappa_0$ is a real constant. By using $(4.5)-(4.7)$, one
can obtain the Weyl anomalous Ward identity obeyed by $I^{\rm ren}_{DS}(h)$.
This can be written in rather explicit form, because of the simple
dependence of $H_h$ and $\varpi_{H_h}$ on $h$. By a somewhat lengthy but
straightforward calculation, one finds
$$
\delta I^{\rm ren}_{DS}(h)=
-{c_{DS}\over 12\pi}\int_\Sigma d^2z \delta\ln h f_h
+{\kappa_0-\kappa_{DS}\over\pi}\delta\int_\Sigma d^2zh^{-1}f_h{}^2,
\eqno(4.15)
$$
where
$$
c_{DS}=-2\tr\big[\big(6(\ad t_0)^2+6\ad t_0+1\big)p_{\goth x}\big],
\eqno(4.16)
$$
$$
\kappa_{DS}=\tr\big(\ad t_{+1}\ad t_{-1}p_{\goth x}\big).
\eqno(4.17)
$$
Choosing $\kappa_0=\kappa_{DS}$ yields a renormalized effective action
$I^{\rm ren}_{DS}(h)$ describing a conformal field theory of central charge
$c_{DS}$. This is precisely the central charge of the DS ghost system
as computed with the methods of hamiltonian reduction and conformal
field theory \ref{14} \footnote{${}^3$}{The odd looking sign of the mid term
in the right hand side of $(4.16)$ is due to the fact that $\goth x$
is negative graded.}. For a generic value of $\kappa_0$, one obtains
a renormalized effective action with a $\int\sqrt h R_h{}^2$ term
yielding a model of induced $2d$ gravity of the same type as that considered
in refs. \ref{18--19}.

It is possible to compute the index of the ghost kinetic operator
$\bar\partial$ in the above framework. One uses the general relation
$(2.32)$ and carries out the calculation using the convenient fiber
metric $H_h$. The result is
$$
\eqalignno{
\ind\bar\partial
=&-{r_{DS}\over 2\pi}\int_\Sigma d^2zf_h
+{1\over\pi}\int_\Sigma d^2z
\tr\big((\ad F_{H_h}+\bar\partial\partial_{H_h}\varpi_{H_h})\varpi_{H_h}\big)
&\cr
=&-\tr\big[\big(2\ad t_0+1\big)p_{\goth x}\big](\ell-1).
&(4.18)\cr}
$$
The dimension of the kernel of $\bar\partial$ is the number of linearly
independent $\gamma$--zero modes. It can be computed as follows.
Recall that to any linearly independent
generator of $\goth g$ of $t_0$ degree $-m<0$ there correspond $d_m$
linearly independent holomorphic sections of $\Ad L$, where $d_m$ is the
dimension of space ${\cal S}^{\rm hol}_{m,0}$ of holomorphic elements
of ${\cal S}_{m,0}$ \ref{16}. Recall also that $d_1=\ell$ and that
$d_m=(2m-1)(\ell-1)$ for $m\geq{3\over 2}$ and $\ell\geq 2$ \ref{17}.
Using these remarks and $(4.3)$, one finds that
$$
\dim\ker\bar\partial=
\dim\goth g_{-1}+{1\over 2}\dim\goth g_{-{1\over 2}}d_{1\over 2}
-\tr\big[\big(2\ad t_0+1\big)p_{\goth x}\big](\ell-1),
\quad \ell\geq 2.
\eqno(4.19)
$$
The dimension of the cokernel of $\bar\partial$ is the number of linearly
independent $\beta$--zero modes. This can be easily computed using
$(4.18)$ and $(4.19)$. One finds
$$
\dim\coker\bar\partial=
\dim\goth g_{-1}+{1\over 2}\dim\goth g_{-{1\over 2}}d_{1\over 2},
\quad \ell\geq 2.
\eqno(4.20)
$$

The above analysis has been carried out for a fixed
$\Ad L_{\goth x}$--special holomorphic structure of the smooth vector
bundle $\Ad L$ characterized by the holomorphic $G$--valued $1$--cocycle
$(4.4)$. One may take such holomorphic structure as a reference one. Let us
now study the family of $\Ad L_{\goth x}$--special holomorphic structure of
$\Ad L$.

Let $R$ be a holomorphic projective connection. Then
$$
A_R={1\over 2}t_{+1}-Rt_{-1}
\eqno(4.21)
$$
is a holomorphic $(1,0)$ connection of $L$. Below, $A_R$ will be used as
background. All fields built using $A_R$ will carry a subscript $R$.

For any Beltrami field $\mu$, consider the holomorphic structure
$\sans S_\mu=(\mu,\ad A^*_R(\mu))$ whose Koszul field $A^*_R(\mu)$ is of
the form
$$
A^*_R(\mu)={1\over 2}\mu t_{+1}-\partial\mu t_0-(\partial^2+R)\mu t_{-1}
\eqno(4.22)
$$
\footnote{${}^4$}{Strictly speaking, the Koszul field is $\ad A^*_R$.
However, in this section, I shall use this name for the field $A^*_R$ itself.}.
It is straightforward to verify that $A^*_R(\mu)$ belongs to
${\cal S}_{0,1,\Ad L}$, so that $A^*_R(\mu)$ is a {\it bona fide}
Koszul field.
A generic holomorphic structure $\sans S=(\mu,\ad A^*_R)$ of $\Ad L$ can be
written in the form
$$
A^*_R=A^*_R(\mu)+a^*,
\eqno(4.23)
$$
where $a^*$ is some element of ${\cal S}_{0,1,\Ad L}$.
Let $\goth S_{DS}$ be the family of all holomorphic structures
$\sans S=(\mu,\ad A^*_R)$ such that $a^*$ is $\goth x$--valued.
Then $\goth S_{DS}\subset\goth S_{\Ad L_{\goth x}}$, {\it i. e.}
$\goth S_{DS}$ consists of $\Ad L_{\goth x}$--special holomorphic structures.
\par\noindent{\it Proof}.
To begin with, one notes that, for the DS bundle, one has
$$
\bar\partial\varpi_H\varpi_H=0,\quad\partial\varpi_H\varpi_H=0.
\eqno(4.24)
$$
The first relation is just $(2.17)$. For a general vector bundle, the second
relation would not be covariant. However, here, because of the specific form
of the cocycle $(4.4)$ and the fact that $\goth x$ is invariant under
$\ad t_0$ and $\ad t_{-1}$, it actually is. $(4.24)$ is shown as follows.
Let $L_0$ be the holomorphic $G$--valued 1--cocycle defined by
$L_{0ab}=\exp(-\ln k_{ab}t_0)$. A generic metric $H$ of $L$ undergoes a Gauss
type factorization of the form
$$
H=KH_0K^\dagger,
\eqno(4.25)
$$
where $K$ is an $\exp\goth x$--valued section of $L\otimes L_0^\vee$ and
$H_0$ is some metric of $L_0$ valued in $\exp\goth k_0$ with $\goth k_0
=\goth q_{-{1\over 2}}\oplus\goth g_0\oplus\goth q_{-{1\over 2}}{}^\dagger$.
Next, pick a basis $\{e_\xi|\xi\in I\}$ of $\goth x$ constituted by
eigenvectors of $\ad t_0$. Then, one has
$$
\eqalignno{
\varpi_H
=&~\Ad K\sum_{\xi,\eta\in I}e_\xi\otimes g(H_0)^{-1}{}_{\xi\eta}\tilde e_\eta
\Ad H_0{}^{-1}\Ad K^{-1},
&\cr
g(H_0)_{\xi\eta}=&~\tilde e_\xi(\Ad H_0{}^{-1}e_\eta),
&(4.26)\cr}
$$
where $\tilde e_\eta=\tr_{\rm ad}(e_\eta^\dagger\hskip1pt\cdot\hskip1pt)$.
{}From this expression, it is not difficult to check the validity of $(4.24)$.
Using $(4.21)$, $(4.23)$ and $(4.24)$ and the fact that $\goth x$ is
invariant under $\ad t_0$ and $\ad t_{-1}$, one finds that $\sans S=
(\mu,\ad A^*_R)$ fulfills $(3.9)$ when $a^*$ is $\goth x$--valued, so that
$\sans S$ is special. \hfill{\it QED}
\par\noindent
Note that $\goth S_{DS}$ is strictly contained in $\goth S_{\Ad L_{\goth x}}$.
For instance, if $\omega^*\in{\cal S}_{0,1}$ and
$\gamma$ is a $(1,0)$ connection of the line bundle $k$, then, setting
$a^*=\omega^*t_0-\gamma\omega^*t_{-1}$, the holomorphic structure
$(\mu,\ad A^*_R)$ defined by $(4.23)$ is special but it is not contained in
$\goth S_{DS}$. In $W$--algebras, $\goth S_{DS}$ is the relevant class
of special holomorphic structures since the constraint on the Wess--Zumino
current is implemented at the lagrangian level by coupling it to a
$\goth x$--valued gauge field, namely $a^*$ \ref{14--15}.

Next, consider the automorphism ghosts $c$ and $M$. Set
$$
M(c,\bar c,\mu)=\big(\partial c+\mu\partial\bar c\big)t_0
+\big(\partial(\partial c+\mu\partial\bar c)
+\partial\mu\partial\bar c\big)t_{-1}.
\eqno(4.27)
$$
One can verify that $M(c,\bar c,\mu)-cA_R$ is a section of $\Ad L$ valued
in $\bigwedge^1(\Lie\Aut_{cL})^\vee$. Using $(3.10)$ and $(3.12)$, one
verifies that $M(c,\bar c,\mu)$ fulfills $(3.11)$.
Write
$$
M=M(c,\bar c,\mu)+m
\eqno(4.28)
$$
with $m$ a section of $\Ad L$ valued in $\bigwedge^1(\Lie\Aut_{cL})^\vee$.
Since $M$ and $M(c,\bar c,\mu)$ both satisfy $(3.11)$, one has
$$
sm=\big(c\partial+\bar c\bar\partial-\ad M(c,\bar c,\mu)\big)m
-{1\over 2}[m,m].
\eqno(4.29)
$$
Now, it is easily checked that
$(c,M)$ fulfills the specialty condition $(3.21)$ if $m$ is
$\goth x$--valued. Such constraint defines a subgroup $\Aut_{cDS}$
of $\Aut_{c\Ad L_{\goth x}}$.
\par\noindent{\rm Proof}.
It follows trivially from $(4.24)$ that $(c,M)$ fulfills the specialty
condition $(3.21)$ once the $\goth x$--valuedness of $m$ is enforced.
Note that $\goth x$--valuedness of $m$ is respected by $(4.29)$,
since $\goth x$ is a subalgebra of $\goth g$ invariant under $\ad t_0$ and
$\ad t_{-1}$. Hence, the constraint defines a subgroup $\Aut_{cDS}$ of
$\Aut_{c\Ad L_{\goth x}}$. \hfill{\it QED}
\par\noindent
Reasoning in the same way as at the end of the previous paragraph, one can
see that $\Aut_{cDS}$ is strictly contained in $\Aut_{c\Ad L_{\goth x}}$.
In $W$--gravity, however, the relevant symmetry group is $\Aut_{cDS}$
since the renormalized matter effective action is invariant only under
$\Aut_{cDS}$ when the background holomorphic structures $\sans S$
are constrained to belong to $\goth S_{DS}$ \ref{14--15}.

Following $(3.15)$, one defines
$$
X_R(C)=cA_R+\bar cA^*_R(\mu)-M(c,\bar c,\mu).
\eqno(4.30)
$$
As suggested by the notation, $X_R(C)$ depends on $c$, $\bar c$ and $\mu$
through the combination $C$ defined in $(3.14)$. In fact
$$
X_R(C)={1\over 2}C t_{+1}-\partial C t_0-(\partial^2+R)C t_{-1}.
\eqno(4.31)
$$
Remarkably, $A^*_R(\mu)$ fulfills $(3.13)$ with $X_A$ replaced by $X_R(C)$
and $X_R(C)$ fulfills $(3.17)$.
It follows from $(3.15)$, $(4.23)$ and $(4.28)$ that
$$
X_R=X_R(C)+x,
\eqno(4.32)
$$
where $x$ is a section of $\Ad L$ valued in $\bigwedge^1(\Lie\Aut_{cL})^\vee$.
Explicitly, from $(3.15)$, $(4.23)$, $(4.28)$ and $(4.30)$, one has
$$
x=\bar ca^*-m.
\eqno(4.33)
$$
Using the fact that $A^*_R$ and $A^*_R(\mu)$ both obey $(3.13)$ with the
appropriate ghost field $X_R$ and that $X_R$ and $X_R(C)$ both obey
$(3.17)$, one finds the relations
$$
sa^*=\big(C\partial_R+\ad X_R(C)\big)a^*
+\big(\bar\partial-\mu\partial_R-\ad A^*_R(\mu)-\ad a^*\big)x,
\eqno(4.34)
$$
$$
sx=\big(C\partial_R+\ad X_R(C)\big)x+{1\over 2}[x,x],
\eqno(4.35)
$$
where $\partial_R=\partial_{A_R}$. If $a^*$ is $\goth x$--valued,
so that the corresponding holomorphic structure is special, then
$(C,X_R)$ fulfills the specialty condition $(3.22)$ if $x$ is
$\goth x$--valued. Note that $(4.34)$ and $(4.35)$ respect
$\goth x$--valuedness.

In the Beltrami--Koszul parametrization, restricting
to holomorphic structures $\sans S\in\goth S_{DS}$,
the DS ghost action reads
$$
S_{DS}(\beta,\beta^\dagger,\gamma,\gamma^\dagger;\mu,\bar\mu,a^*, a^{*\dagger})
={1\over\pi}\int_\Sigma d^2z \tr_{\rm ad}\big[\beta
\big(\bar\partial-\mu\partial_R-\ad A^*_R(\mu)-\ad a^*\big)\gamma\big]
+{\rm c.~c.},
\eqno(4.36)
$$
where $a^*$ is $\goth x$--valued.
Using this expression, it is straightforward to compute the classical
energy--momentum tensor $T_{DS}(\beta,\gamma)$. One has
$$
\eqalignno{
T_{DS}(\beta,\gamma)=&~\pi{\delta S_{DS}\over\delta\mu}
(\beta,\beta^\dagger,\gamma,\gamma^\dagger;0,0,0,0)
=\tr_{\rm ad}\big(\partial_R\gamma\beta+D_R[\gamma,\beta]\big),
&\cr
D_R=&~{1\over 2}t_{+1}+t_0\partial-t_{-1}(\partial^2+R).
&(4.37)\cr}
$$
Similarly, one can compute the classical gauge current $J_{DS}(\beta,\gamma)$.
One finds
$$
J_{DS}(\beta,\gamma)=\pi{\delta S_{DS}\over\delta a^*}
(\beta,\beta^\dagger,\gamma,\gamma^\dagger;0,0,0,0)=[\gamma,\beta].
\eqno(4.38)
$$
Note that $T_{DS}(\beta,\gamma)$ contains a second derivative improvement term
$\tr_{\rm ad}\big(D_RJ_{DS}(\beta,\gamma)\big)$, a common feature in
$W$--algebras. Note also that $J_{DS}(\beta,\gamma)$ is valued in $\goth g/
[\goth x,\goth x^\perp]$ since $\beta$ is $\goth g/\goth x^\perp$--valued and
$\gamma$ is $\goth x$--valued.

In the above geometrical formulation, I have not defined a notion of
stability for special holomorphic structures $\sans S=(\mu,\ad A^*_R)
\in\goth S_{DS}$ with a fixed Beltrami field $\mu$. In the analysis below,
it will be  assumed that the holomorphic structure on $\Sigma$ defined by
$\mu$ is generic in the sense that $d_{1\over 2}=0,1$ depending on whether
the spinor structure is even or odd, respectively. Now, no structure
$\sans S\in\goth S_{DS}$ is stable in the customary sense. Indeed,
the space ${\cal S}^{\rm hol}_{0,0\Ad L;\ssans S}$ of holomorphic
elements in ${\cal S}_{0,0\Ad L:\ssans S}$ is non trivial, while, for
stable structures, ${\cal S}^{\rm hol}_{0,0\Ad L;\ssans S}$ must vanish
\ref{17}. In physical terms, ${\cal S}^{\rm hol}_{0,0\Ad L;\ssans S}$ is
the space of the holomorphic infinitesimal gauge transformations of
$\Ad L_{\ssans S}$ and, for stable structures $\sans S$, has minimal
dimension. Here, the relevant symmetry group is the DS gauge group
$\Gau_{cDS}$, which is the gauge subgroup of $\Aut_{cDS}$
\footnote{${}^5$}{In geometrical terms, an element $\alpha$ of the
automorphism group $\Aut_{cE}$ of a vector bundle $E$ is a
gauge transformation if the induced diffeomorphism $f_\alpha=\id_\Sigma$.}.
So, one may define stability as follows. $\sans S$ is said stable
if the space ${\cal S}^{\rm hol}_{0,0\Ad L_{\goth x};\ssans S}$ of holomorphic
$\goth x$--valued elements in ${\cal S}_{0,0\Ad L;\ssans S}$ has minimal
dimension. Let us denote by $\goth S^{\rm stab}_{DS}$ the subspace of
$\goth S_{DS}$ of all stable holomorphic structures $\sans S$ of
$\goth S_{DS}$. Clearly, $\goth S^{\rm stab}_{DS}$ is preserved by
$\Aut_{cDS}$. Non stable holomorphic structures must satisfy in the
Beltrami--Koszul parametrization certain linear conditions. They thus
span a submanifold of $\goth S_{DS}$ of finite codimension. Hence,
$\goth S^{\rm stab}_{DS}$ is dense in $\goth S_{DS}$.

In $W$--gravity there are two geometrical structures of crucial
importance in analogy to string theory. The first is
the holomorphic subgroup $\Gau^{\rm hol}_{cDS;\ssans S}$ of
the DS gauge group $\Gau_{cDS}$ for any stable holomorphic structure
$\sans S\in\goth S^{\rm stab}_{DS}$. The second is DS Teichmueller space
$\Teich_{DS}=\goth S^{\rm stab}_{DS}/\Gau_{cDS}$ of stable
holomorphic structures $\sans S\in\goth S^{\rm stab}_{DS}$
modulo $\Gau_{cDS}$. Their dimensions can be computed. By direct calculation,
one finds that
$$
\dim\Gau^{\rm hol}_{cDS;\ssans S}
=\dim\goth g_{-1}+n_*d_{1\over 2}
-\tr\big[\big(2\ad t_0+1\big)p_{\goth x}\big](\ell-1),
\quad \ell\geq 2,
\eqno(4.39)
$$
where $n_*=\min_{x\in\goth p_{-{1\over 2}}}
\dim\ker\ad x|_{\goth p_{-{1\over 2}}}$.
Clearly, $n_*$ depends on $\goth s$ and $n_*\geq 1$.
Using the index relation $(4.18)$ and $(4.39)$, one finds
that
$$
\dim\Teich_{DS}=
\dim\goth g_{-1}+n_*d_{1\over 2},
\quad \ell\geq 2.
\eqno(4.40)
$$
The calculation of these numbers is one of the main results of this paper.
\par\noindent{\it Proof}.
Consider the holomorphic structure $\sans S_\mu=(\mu,\ad A^*_R(\mu))$
defined earlier. It is not difficult to check that the intertwiner
$V_{\ssans S_\mu}$ of $\sans S_\mu$ is given by
$\exp(-\ln\partial z_{\ssans S_\mu}t_0)\exp(\partial(\partial
z_{\ssans S_\mu})^{-1}t_{-1})$ and that $L_{\ssans S_\mu ab}=
\exp(-\ln k_{\ssans S_\mu ab}t_0)\exp(\partial_a k_{\ssans S_\mu ab}{}^{-1}
t_{-1})$.
Note that this $1$--cocycle is of the DS form $(4.4)$.
Hence, choosing the reference holomorphic structure of $\Ad L$, so that
the induced holomorphic structure on $\Sigma$ is generic in the sense
stated above, one can assume that $\mu=0$ without loss of generality.
The holomorphic structures $\sans S\in\goth S_{DS}$, in which one is
interested, are therefore of the form $(0,a^*)$ with $a^*$ an
$\goth x$--valued element of ${\cal S}_{0,1,\Ad L}$.
Let $\Theta$ be a section of $\bar k^{\otimes\bar w}\otimes\Ad L$. One can
decompose $\Theta$ as follows
$$
\Theta=\sum_{m\in\Bbb Z/2,|m|\leq j_*}\Theta^{(m)}\quad{\rm with}\quad
[\ad t_0,\Theta^{(m)}]=m\Theta^{(m)},
\eqno(4.41)
$$
where $j_*$ is the highest eigenvalue of $\ad t_0$. Applying theorems
$3.2$ and $3.3$ of ref. \ref{16}, one can easily show the following. If
$\Theta^{(m)}=0$ for $p<m\leq j_*$ with $-j_*\leq p<j_*$, then
$\Theta^{(p)}$ is a section of $k^{\otimes-p}\otimes\bar k^{\otimes\bar w}
\otimes\goth g_p$. Pick a holomorphic projective connection $R$.
For any section $\theta$ of
$k^{\otimes-p}\otimes\bar k^{\otimes\bar w}\otimes\goth g_p$ with
$-j_*\leq p\leq-{1\over 2}$, there exists a section $T_R(\theta)$ of
$\bar k^{\otimes\bar w}\otimes\Ad L$ such that
$$
T_R(\theta)^{(m)}=0,\quad{\rm for}~p<m\leq j_*,\quad
T_R(\theta)^{(p)}=\theta.
\eqno(4.42)
$$
Further, when $\bar w=0$, one has
$$
\bar\partial T_R(\theta)=T_R(\bar\partial\theta).
\eqno(4.43)
$$
Consider the equation
$$
\big(\bar\partial-\ad a^*\big)\eta=0
\eqno(4.44)
$$
with $\eta$ an $\goth x$--valued section of $\Ad L$. The space of solution
of this equation is precisely $\ker\bar\partial_{\ssans S}\cong
\Lie\Gau^{\rm hol}_{cDS;\ssans S}$. Now, set $\eta_0=\eta$. Then, by
$(4.3)$, $\eta_0{}^{(m)}=0$ for $m>-{1\over 2}$, so that
$\eta_0{}^{(-{1\over 2})}$ is a section of
$k^{\otimes{1\over 2}}\otimes\goth g_{-{1\over 2}}$, as recalled above.
It follows from $(4.44)$ and the fact that $a^{*(m)}=0$ for
$m>-{1\over 2}$ that $\bar\partial\eta_0{}^{(-{1\over 2})}=0$
by grading reasons. There are
$d_{1\over 2}\dim\goth g_{-{1\over 2}}/2$ linearly independent such
$\eta_0{}^{(-{1\over 2})}$. Define $\eta_1=\eta_0
-T_R\big(\eta_0{}^{(-{1\over 2})}\big)$. By $(4.42)$, one has that
$\eta_1{}^{(m)}=0$ for $m>-1$, so that $\eta_1{}^{(-1)}$ is a section of
$k\otimes\goth g_1$. By $(4.43)$, the holomorphicity of
$\eta_0{}^{(-{1\over 2})}$, $t_0$--grading reasons and $(4.44)$ one has
further that $\bar\partial\eta_1{}^{(-1)}=
[a^{*(-{1\over 2})},\eta_0{}^{(-{1\over 2})}]$. The general solution
of this equation, if it exists, is a linear inhomogeneous function of
$d_1\dim\goth g_{-1}$ complex parameters since $\eta_1{}^{(-1)}$ is
determined up to the addition of an arbitrary section $\zeta^{(-1)}$
of $k\otimes\goth g_1$ such that $\bar\partial\zeta^{(-1)}=0$.
A solution exists
provided the integrability condition $\int_\Sigma d^2 z
[a^{*(-{1\over 2})},\eta_0{}^{(-{1\over 2})}]=0$ is satisfied.
Since $d_{1\over 2}=0,1$, $\eta_0{}^{(-{1\over 2})}$ is of the form
$\sigma x^{(-{1\over 2})}$, where $\sigma$ is a holomorphic section of
$k^{\otimes{1\over 2}}$ such that $\sigma\not=0$ if
$d_{1\over 2}=1$ and $x^{(-{1\over 2})}\in\goth p_{-{1\over 2}}$.
Hence, the integrability condition reduces into
$[\int_\Sigma d^2 z\sigma a^{*(-{1\over 2})},x^{(-{1\over 2})}]=0$.
If $a^*$ is to represent a stable holomorphic structure, this must be a
condition constraining $x^{(-{1\over 2})}$ only. From here, it is easy to
see that, for a stable holomorphic structure, the space of allowed
$\eta_0{}^{(-{1\over 2})}$ has dimension $n_*d_{1\over 2}$. Next,
define $\eta_2=\eta_1-T_R\big(\eta_1{}^{(-1)}\big)$.
By $(4.42)$, one has that $\eta_2{}^{(m)}=0$ for $m>-{3\over 2}$,
so that $\eta_2{}^{(-{3\over 2})}$ is a section of
$k^{\otimes{3\over 2}}\otimes\goth g_{-{3\over 2}}$.
By $(4.43)$, one has further that $\bar\partial\eta_2{}^{(-{3\over 2})}=
\big(\bar\partial\eta_1-T_R\big(\bar\partial\eta_1{}^{(-1)}\big)
\big)^{(-{3\over 2})}$. The general solution
of this equation always exists and is a linear inhomogeneous function of
$d_{3\over 2}\dim\goth g_{-{3\over 2}}$ complex parameters,
since $\eta_2{}^{(-{3\over 2})}$ is determined up to the addition of an
arbitrary section $\zeta^{(-{3\over 2})}$
of $k^{\otimes{3\over 2}}\otimes\goth g_{-{3\over 2}}$
such that $\bar\partial\zeta^{(-{3\over 2})}=0$. The procedure
can now be iterated. At the $p$--th step one defines a section
$\eta_p{}^{(-{p+1\over 2})}$ of
$k^{\otimes{p+1\over 2}}\otimes\goth g_{-{p+1\over 2}}$ satisfying an
equation whose general solution is a linear inhomogeneous function of
$d_{p+1\over 2}\dim\goth g_{-{p+1\over 2}}$ complex parameters. In conclusion,
for a stable structure, $\dim\ker\bar\partial_{\ssans S}=
n_*d_{1\over 2}+\sum_{p\geq 1}d_{p+1\over 2}\dim\goth g_{-{p+1\over 2}}$.
Using that $d_1=\ell$ and that
$d_m=(2m-1)(\ell-1)$ for $m\geq{3\over 2}$ and $\ell\geq 2$
\ref{17} and the remark just below $(4.44)$, one obtains $(4.39)$ readily.
To compute $\dim\Teich_{DS}$, one notes $\dim\Teich_{DS}
=\dim\Gau^{\rm hol}_{cDS;\ssans S}-\ind\bar\partial_{\ssans S}$ by a reasoning
analogous to that used to compute the dimension of the ordinary Teichmueller
space in string theory. Then, $(4.40)$ follows immediately from
the index relation $(4.18)$ and $(4.39)$.
\hfill{\it QED}
\vskip.6cm
\par\noindent
{\bf Acknowledgements.} I wish to voice my gratitude to M. Matone,
E. Aldrovandi, G. Velo and expecially M. Bauer and R. Stora for helpful
discussions.
\vskip.6cm
\centerline{\bf REFERENCES}
\def\ref#1{\lbrack #1\rbrack}
\def\NP#1{Nucl.~Phys.~{\bf #1}}
\def\PL#1{Phys.~Lett.~{\bf #1}}

\def\CMP#1{Commun.~Math.~Phys.~{\bf #1}}
\def\PR#1{Phys.~Rev.~{\bf #1}}
\def\PRL#1{Phys.~Rev.~Lett.~{\bf #1}}

\def\PREP#1{Phys.~Rep.~{\bf #1}}

\def\JP#1{J.~Phys.~{\bf #1}}
\def\CQG#1{Class.~Quantum~Grav.~{\bf #1}}
\vskip.4cm
\par\noindent

\item{\ref{1}}
S. W. Hawking, \CMP{55} (1977) 133.

\item{\ref{2}}
N. K. Nielsen, H. Roemer and B. Schroer, \NP{B136} (1978) 445.

\item{\ref{3}}
J. Singe, in ``{\it Relativity: the General Theory}", North Holland,
Amsterdam (1960).

\item{\ref{4}}
J. S. Dowker and R. Critchley, \PR{D13} (1976) 3224.

\item{\ref{5}}
J. S. Dowker, \JP{A11} (1978) 347.

\item{\ref{6}}
O. Alvarez, \NP{B216} (1983) 125; \NP{B238} (1984) 61.

\item{\ref{7}}
K. Fujikawa, \PRL{42} (1979) 1195; Phys. Rev. {\bf D21} (1980) 2848;
Phys. Rev. {\bf D22} (1980) 1499(E); Phys. Rev. Lett. {\bf 44} (1980) 1733;
Phys. Rev. {\bf D23} (1981) 2262.

\item{\ref{8}}
R. T. Seeley, Am. Math. Soc. Proc. Symp. Pure Math. {\bf 10} (1967)
288.

\item{\ref{9}}
P. B. Gilkey, Proc. Symp. Pure Math. {\bf 27} (1973); J. Diff. Geom.
{\bf 10} (1975) 601.

\item{\ref{10}}
B. De Witt, in ``{\it Dynamical Theory of Groups and Fields}",
Gordon and Breach (1965).

\item{\ref{11}}
M. Knecht, S. Lazzarini and R. Stora, \PL{B273} (1991) 63.

\item{\ref{12}}
R. Zucchini, \CQG{11} (1994) 1697.

\item{\ref{13}}
J. De Boer and J. Goeree, \NP{B401} (1993) 369.

\item{\ref{14}}
L. Feh\'er, L O'Raifertaigh, P. Ruelle, I. Tsutsui and A. Wipf,
\PREP{222} no. 1, (1992) 1.

\item{\ref{15}}
R. Zucchini, in preparation.

\item{\ref{16}}
R. Zucchini, \PL{B323} (1994) 322; preprint hep-th/9403036 to appear in
J. Geom. Phys. and references therein.

\item{\ref{17}}
R. Gunning, {\it Lectures on Riemann Surfaces}, Princeton University
Press (1966);
R. Gunning, {\it Lectures on Vector Bundles on Riemann Surfaces}, Princeton
University Press (1967).

\item{\ref{18}}
H. Kawai and R. Nakayama, \PL{B306} (1993) 224.

\item{\ref{19}}
S. Ichinose, N. Tsuda and T. Yukawa, preprint hep-th/9502101.

\item{\ref{20}}
N. J. Hitchin, Topology {\bf 31} no. 3, (1992) 449.

\item{\ref{21}}
S. Govindarajan and T. Jayaraman, \PL{B345} (1995) 211.

\item{\ref{22}}
K. K. Uhlenbeck and S. T. Yau, Comm. Pure and Appl. Math. {\bf 39-S}
(1986) 257

\item{\ref{23}}
C. T. Simpson, J. Am. Math. Soc. {\bf 1} no. 4, (1988) 867.

\item{\ref{24}}
V. G. Drinfeld and V. V. Sokolov, J. Sov. Math. {\bf 30} (1985) 1975.

\bye